\documentstyle[epsfig]{mn}
\newcommand{\etal}{{et al.}}

\begin{document}
\title[Jets, lobe size \& spectral index]{Asymmetry of jets, lobe size
and spectral index in radio galaxies and quasars}
\author[J. Dennett-Thorpe et al.]
   {J.Dennett-Thorpe$^{1,4}$, A.~H. Bridle$^2$, R.~A. Laing$^3$, P.~A.~G. Scheuer$^1$\\
     $^1$ Mullard Radio Astronomy Observatory, 
     Cavendish Laboratory, Madingley Rd, Cambridge, UK \\ 
     $^2$ National Radio Astronomy Observatory, 520 Edgemont Rd, Charlottesville, VA, USA\\
     $^3$ Royal Greenwich Observatories, Madingley Rd, Cambridge. UK \\
     $^4$ Observat{\'o}rio Astron{\'o}mico de Lisboa, Tapada da Ajuda, 1300 Lisbon, Portugal\\
}

\date{Received }
\maketitle

\begin{abstract}
  In this paper we investigate the correlations between spectral index,
  jet side and extent of the radio lobes for a sample of nearby FRII
  radio galaxies.  In Dennett-Thorpe \etal\ (1997) we studied a sample
  of quasars and found that the high surface brightness regions had
  flatter spectra on the jet side (explicable as a result of Doppler
  beaming) whilst the extended regions had spectral asymmetries
  dependent on lobe length.  Unified schemes predict that asymmetries
  due to beaming will be much smaller in narrow-line radio galaxies than
  in quasars, and we therefore chose to investigate in a similar
  fashion, a sample of radio galaxies with detected jets.  We find that
  spectral asymmetries in these objects are uncorrelated with jet
  sidedness at all brightness levels, but depend on relative lobe
  volume.  Our results are not in conflict with unified schemes, but
  suggest that the differences between the two samples are due primarily
  to power or redshift, rather than to orientation. We also show
  directly that hotspot spectra steepen as a function of radio power or
  redshift. Whilst a shift in observed frequency due to the redshift may
  account for some of the steepening, it cannot account for all of it, and
  a dependence on radio power is required.
\end{abstract}


\section{Introduction}

Several strong correlations have been found between asymmetries in
powerful extragalactic radio sources of Fanaroff \& Riley's (1974)
\nocite{fan74} class II (hereafter FRII):
\begin{enumerate}
\item In a sample of sources with strong, one-sided jets (predominantly
quasars), the jet side depolarizes less rapidly with increasing wavelength
(Laing 1988; Garrington \etal\ 1988).
\item The jet side also has a flatter spectrum in these objects 
(Garrington, Conway \& Leahy 1991). 
\item In a sample of sources selected without reference to the prominence of
their jets (mostly radio galaxies), the more depolarized lobe has a steeper 
spectrum (Liu and Pooley 1991a, 1991b).
\item In sources with weak or undetected jets (all radio galaxies) the shorter
lobe depolarizes faster (Pedelty \etal\ 1989a, b; Laing 1996) 
\item More extended narrow-line emission is associated with the shorter lobe in
radio galaxies (McCarthy \etal\ 1991).
\end{enumerate}
The Laing-Garrington depolarization--jet side correlation is most easily
explained as an orientation effect. If the jets are relativistic, the
nearer one
appears brighter as a result of Doppler boosting; the nearer lobe is seen
through less magnetoionic material and therefore shows less depolarization.
The association of depolarization and enhanced line emission with the shorter
lobe is more likely to result from intrinsic or environmental effects such as a
variations in external density or motion of the host galaxy (Gopal-Krishna
\etal\ 1996).  This leads to a problem if the spectral-index correlations 
have a common origin in all objects:  
there is an unbroken chain of reasoning which implies that apparently 
intrinsic properties are correlated
with those believed to be orientation--dependent. In Dennett-Thorpe 
\etal\ (1997, hereafter Paper 1) we presented evidence which resolved this paradox for
quasars: the spectrum of the regions of highest brightness, in and
around hotspots, is indeed flatter on the jet side, but that of the
low-brightness emission is flatter in the longer lobe.

The results of Paper 1 indicate the presence of a mixture of
environmental/intrinsic and orientation effects. In the present paper
we have attempted to isolate the former by investigating a sample in
which Doppler beaming should be less important.  According to the
simplest variants of unified schemes (e.g. Barthel
1989),\nocite{bar89} the quasar jet axes lie within $\approx45^\circ$
of the line of sight.  Their side-on counterparts are the narrow-line
radio galaxies (NLRG). The ideal comparison sample would consist of
NLRG comparable in luminosity and redshift to the quasars observed in
Paper 1, but the frequency of detection of jets in these objects in
existing observations is very low (Fernini \etal\ 1993;
1998)\nocite{fer93,fer97} and we cannot investigate correlations with
jet sidedness.  By contrast, the detection rate of jets achieved for
{\it nearby} FRII radio galaxies is high, and we have  selected a
sub-sample of these sources with detected jets.  The sample consists
mainly of NLRG; these should be closer to the plane of the sky than
the quasars, at least on average (note that the need to detect at
least one jet introduces an orientation-dependent bias to the
selection). It also includes two broad-line radio galaxies (BLRG)
which (by the precepts of unfied schemes) should have jets at fairly small angles to the line of sight and
might therefore be expected to behave more like the quasars.  The
difference in characteristic luminosity and redshift between the two
samples is an important but unavoidable complication.

In section 2, we describe the sample selection, observations and derivation of
spectral indices.  Spectral asymmetries and their relation to other jet
sidedness and lobe volume ratio are discussed in Section 3 and the spectral 
differences between the quasar and galaxy samples are  considered in Section 4.
Orientation-dependent asymmetries are considered briefly in 
Section 5, and Section 6 summarizes our conclusions.

We assume throughout that $H_0$ = 50~kms$^{-1}$Mpc$^{-1}$ and $q_0$ = 0.5.

\nocite{lai88,gar88,gar91a} 
\nocite{mcc91,ped89a,ped89b} 
\nocite{gop96,liu91b,liu91c}
\nocite{den97}
\nocite{lai96-contrib}

\section{Sample Selection and Observations}

The observed sources were selected from the sample defined by Black
\etal\ (1992), which is a subset of the 3C catalogue with $P_{178} >$
1.5 $\times$ 10$^{25}$WHz$^{-1}$sr$^{-1}$ and $z <$ 0.15.  Sensitive
high-resolution images of all of the sample members are available from
the work of Black \etal\ (1992)\nocite{bla92a}, Leahy \etal\ (1997)
\nocite{lea97} and the other references listed by Black \etal\ As a
result, the jet detection rate is $\approx$60\% and the sample is
therefore well suited to the purposes of this study.  We selected
sources with detected jets having a largest angular size
$<$300~arcsec, to ensure adequate sampling of the low-brightness
emission. Known FRI sources had already been excluded by Black \etal\,
but we also omitted three sources with intermediate FRI/II or grossly
distorted structure (3C15, 424 and 433).  We were able to obtain
suitable data for a representative sub-sample of 10 out of the
remaining 15 sources.  Those rejected were: 3C111, 277.3 and 303 (for
which archival data proved to be inadequate); 3C184.1 (whose jet had
not been detected when the sample was selected) and 3C353 (the subject
of a detailed study by Swain (1996)). \nocite{swa96}

Unfortunately the sample members turn out to be rather symmetrical, so that
effects of differences in lobe length cannot be expected to stand out as
clearly as we should like.  Jets were detected only on one side of the core 
in 8/10 sources, but we also use the term ``jet side'' to refer to the 
lobe with the brighter jet in the other 2 cases.  This is ambiguous
only for 3C452.

A summary of the observations is given in Table~\ref{tab:obs}. For this work a
well-covered uv-plane is needed, preferably matched at two frequencies in the
sense that the baselines in wavelengths are identical. Each source was observed
at frequencies of approximately 1.4~GHz and either 4.9 or 8.4~GHz, depending
on the availability of suitable data from previous observations (precise
frequencies are given later). Where necessary, we undertook extra VLA
observations at the higher frequency to improve the coverage at large spatial
scales. At 1.4~GHz we combined archive data and new observations to create a
data-set which substantially overlapped in the uv-plane with the higher
frequency (we were unable to match the arrays by individual baselines at
8.4~GHz, as the VLA configurations do not scale appropriately). New
observations are marked with an asterisk, and were taken on 1995 June 18 (B
array), 1995 November 26 (C array) and 1995 April 11 (D array).  References
are given in Table~\ref{tab:obs} to the sources of the archive data. In
most cases, the data were recalibrated and mapped for the present project.
The exceptions were 3C382, where the low-frequency image is from Leahy \&
Perley (1991)\nocite{lea91} and 3C388 (Roettiger \etal\ 1994) and 3C405
(Dreher \etal\ 1987; Carilli \etal\ 1991), where published maps were used
at both frequencies. \nocite{roe94,dre87,car91}

\begin{table}
\caption{Observations}
\label{tab:obs}
\begin{tabular}{llll}
Source & Frequency & Array & Observations\\
3C135 &1.4& BC & $\star$\\
      &8.4& BCD &L97\\
3C192 &1.4& BC &RAL,$\star$\\
      &4.9& CD &RAL,$\star$\\
3C223.1&1.4& BC &SS,$\star$\\
      &8.4& ABCD&B92,$\star$\\
3C285 &1.4& ABC & vBD\\
      &4.9& ABCD & vBD, $\star$\\
3C382 &1.4& BC &LP91\\
      &8.4& BCD &B92\\
3C388 &1.4& $\dagger$ &R94\\
      &4.9& $\dagger$ &R94\\
3C390.3&1.4& ABC & LP95\\
      &8.4& CD & LP95, $\star$\\
3C403 &1.4& BC & $\star$\\
      &8.4& CD &B92, $\star$\\
3C405 &1.4& ABD &D87\\
      &4.9& ABCD &D87\\
3C452 &1.4& BC & RAL\\
      &8.4& ABCD & B92, $\star$\\
\end{tabular}
\parbox{\linewidth}{$\star$ marks new observations undertaken for this
  project.}

\parbox{\linewidth}{$\dagger$ non-standard, but matched, arrays. See
  Roettiger \etal\ (1994)}
\parbox{\linewidth}{References: B92 Black \etal\ (1992); 
D87 Dreher \etal\ (1987); LP91 Leahy \& 
  Perley (1991);LP95 Leahy \& 
  Perley (1995)\nocite{lea95}; L97 Leahy \etal\
  (1997); R94 Roettiger \etal\ (1994); vBD van
Breugel \& Dey (1993)\nocite{bre93}; RAL, SS R.A. Laing, S.R. Spangler 
 (unpublished).}
\end{table}

\begin{table*}
\caption{Image properties}
\label{tab:images}
\begin{tabular}{lllllll}
Source & beam&$\nu_1$(MHz) & noise & contour& $\nu_2$(MHz)& noise\\
       & (arcsec)&         & mJy/beam& factor &             & mJy/beam\\
3C135  & 4.65 & 1450 & 0.17 & 2.02 & 8350 & 0.09\\
3C192  & 4.35 & 1418 & 0.18 & 1.82 & 4910 & 0.05\\
3C223.1& 10.0 & 1440 & 0.32 & 2.67 & 8350 & 0.13\\
3C285  & 5.50 & 1506 & 0.12 & 1.86 & 4860 & 0.09\\
3C382  & 4.75 & 1477 & 0.13 & 2.00 & 8350 & 0.10\\
3C388  & 1.45 & 1465 & 0.06 & 1.89 & 4885 & 0.04\\
3C390.3& 4.85 & 1533 & 0.23 & 2.12 & 8350 & 0.20\\
3C403  & 4.50 & 1450 & 0.15 & 1.98 & 8350 & 0.11\\
3C405  & 1.33 & 1446 & 6.36 & 2.13 & 4525 & 0.76\\
3C452  & 5.25 & 1403 & 0.30 & 1.68 & 8350 & 0.09\\
\end{tabular}
\end{table*}

The data were reduced with the NRAO {\sc aips} software, using
standard procedures and phase-only self-calibration.  Limits and
tapers were applied to the data in the uv-plane to create data sets
with approximately matched baselines and beam sizes. After extensive
testing of different imaging algorithms, we decided to use the {\sc
clean}-based routines {\sc mx} and {\sc imagr}, in preference to the
maximum entropy-based alternative {\sc vtess} (the one exception is
the low-frequency image of 3C382, where we used the published
{\sc clean-mem} hybrid image from Leahy \& Perley 1991). This was done in
order to ensure the reliability and reproducibility of the high and
low surface-brightness regions; by contrast, convergence of
maximum-entropy algorithms for these images was very sensitive to the
input value of the noise, and was possible only after several
attempts. The images were cleaned down to the noise level, to ensure a
good representation of the flux density even in the lowest surface
brightness regions. The fraction of the flux density represented by
{\sc clean} components in any part of the image is easily assessed
using the {\sc aips} software. The resolution used was usually
determined by the requirements that $>95\%$ of the flux density of the
final image was represented by {\sc clean} components and that the
resulting image remained free of obvious imaging artifacts. In sources
with extended low surface brightness regions, further tests were
performed to ensure that $>95\%$ of the flux density of these regions
alone was also represented by {\sc clean} components.

\begin{figure*}
  \centerline{
    \epsfig{figure=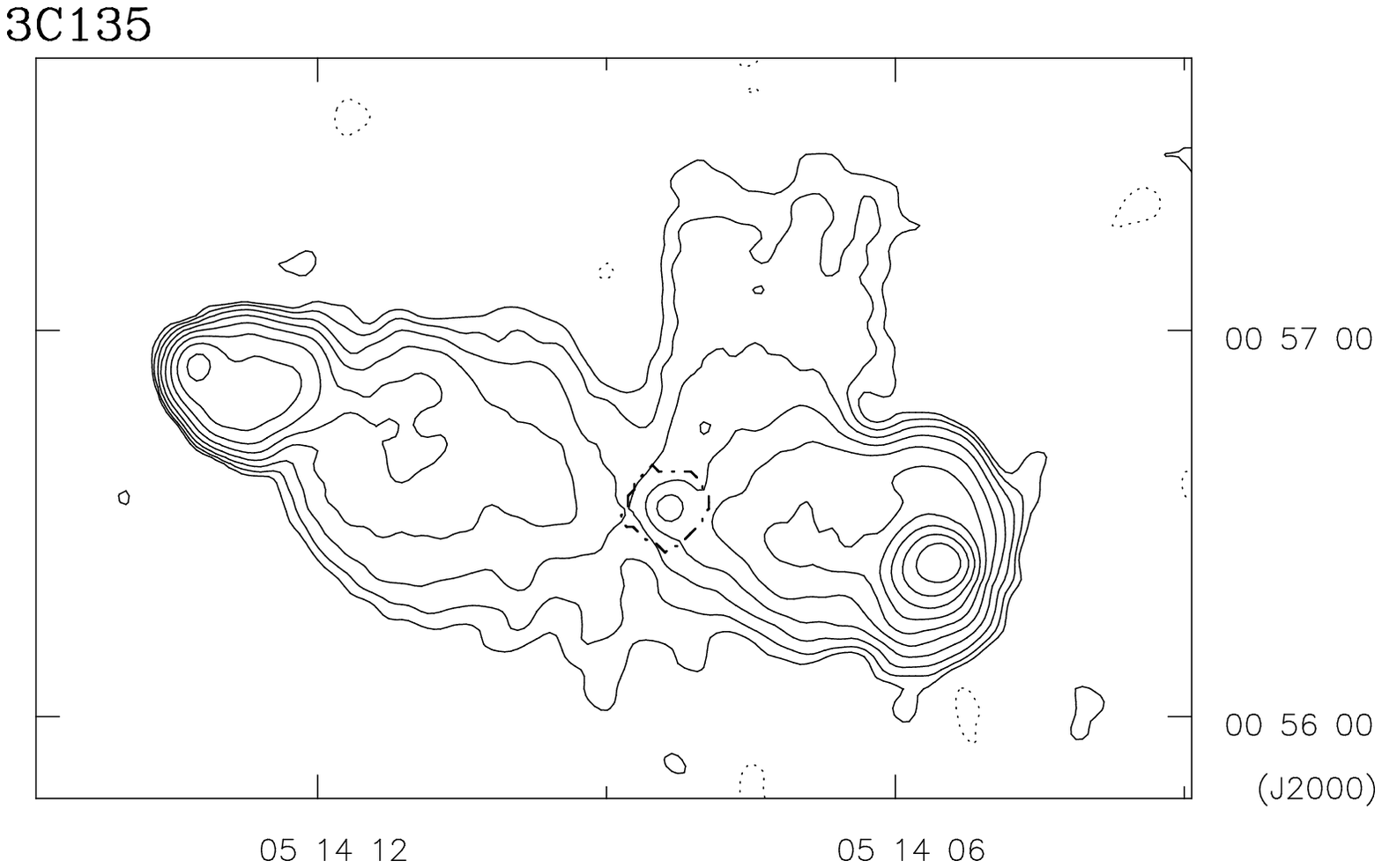,width=6.2cm, angle=0}
    \epsfig{figure=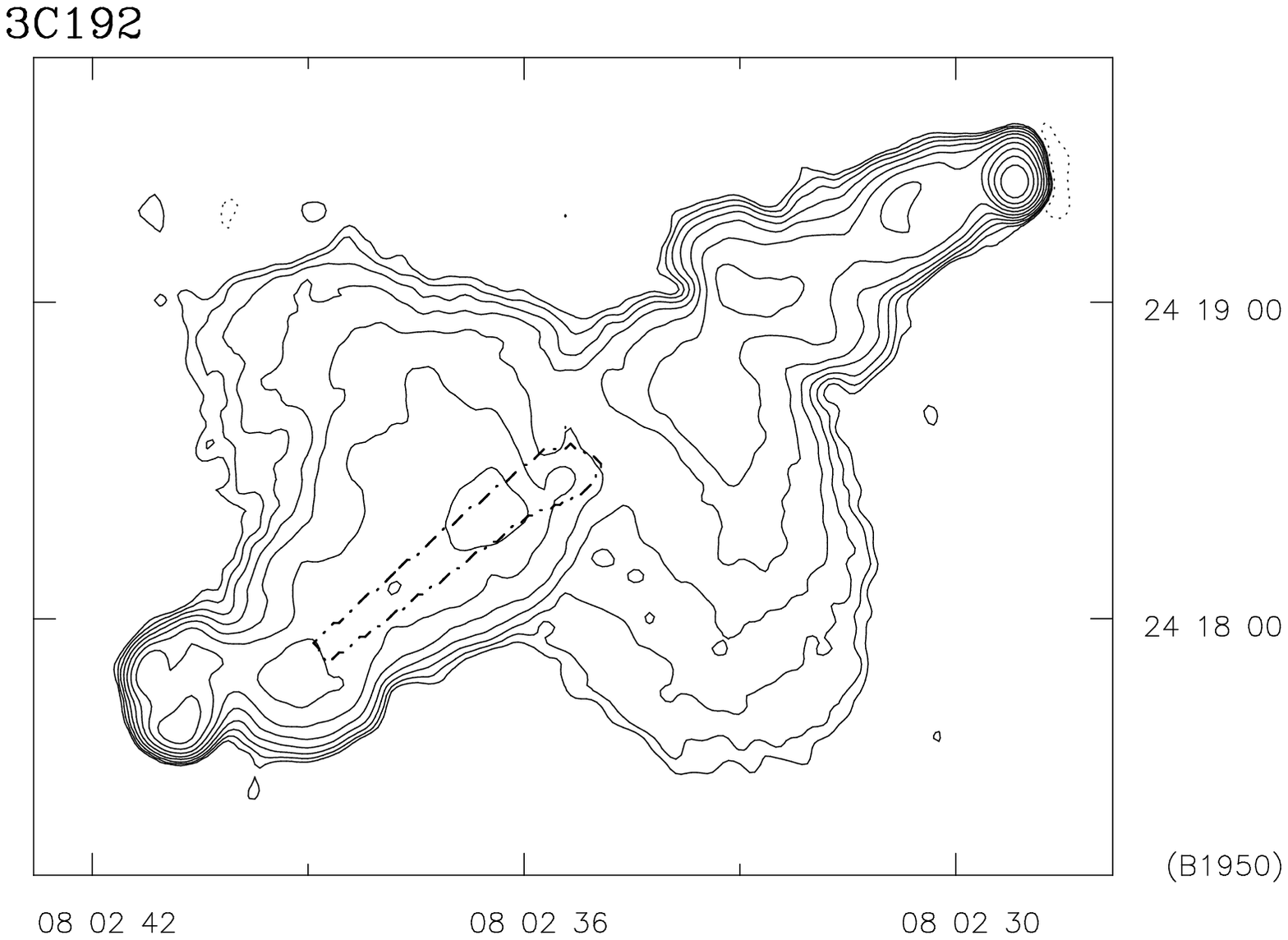,width=5.2cm, angle=0}} 
  \centerline{
    \epsfig{figure=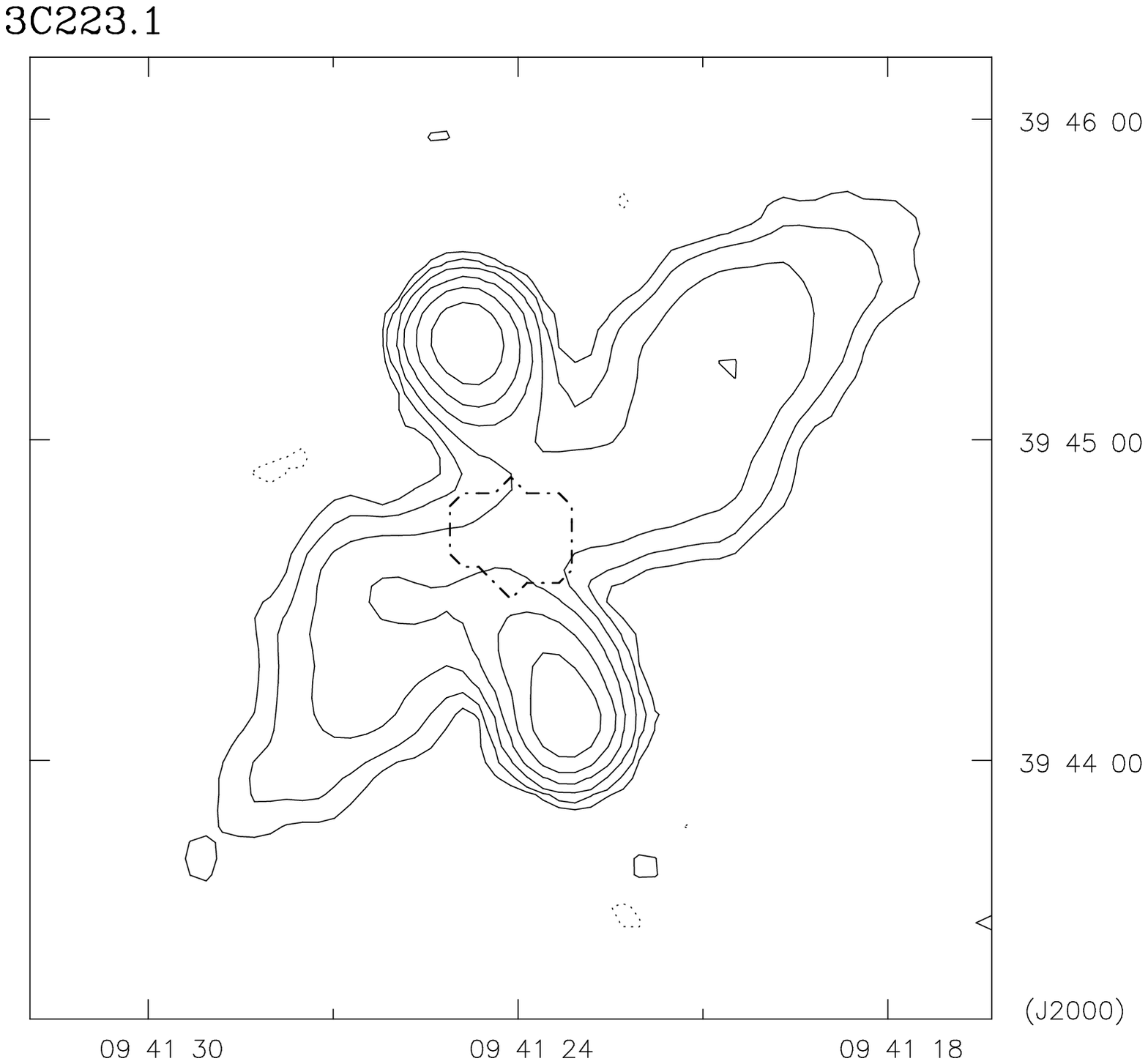,width=5.0cm, angle=0}
    \epsfig{figure=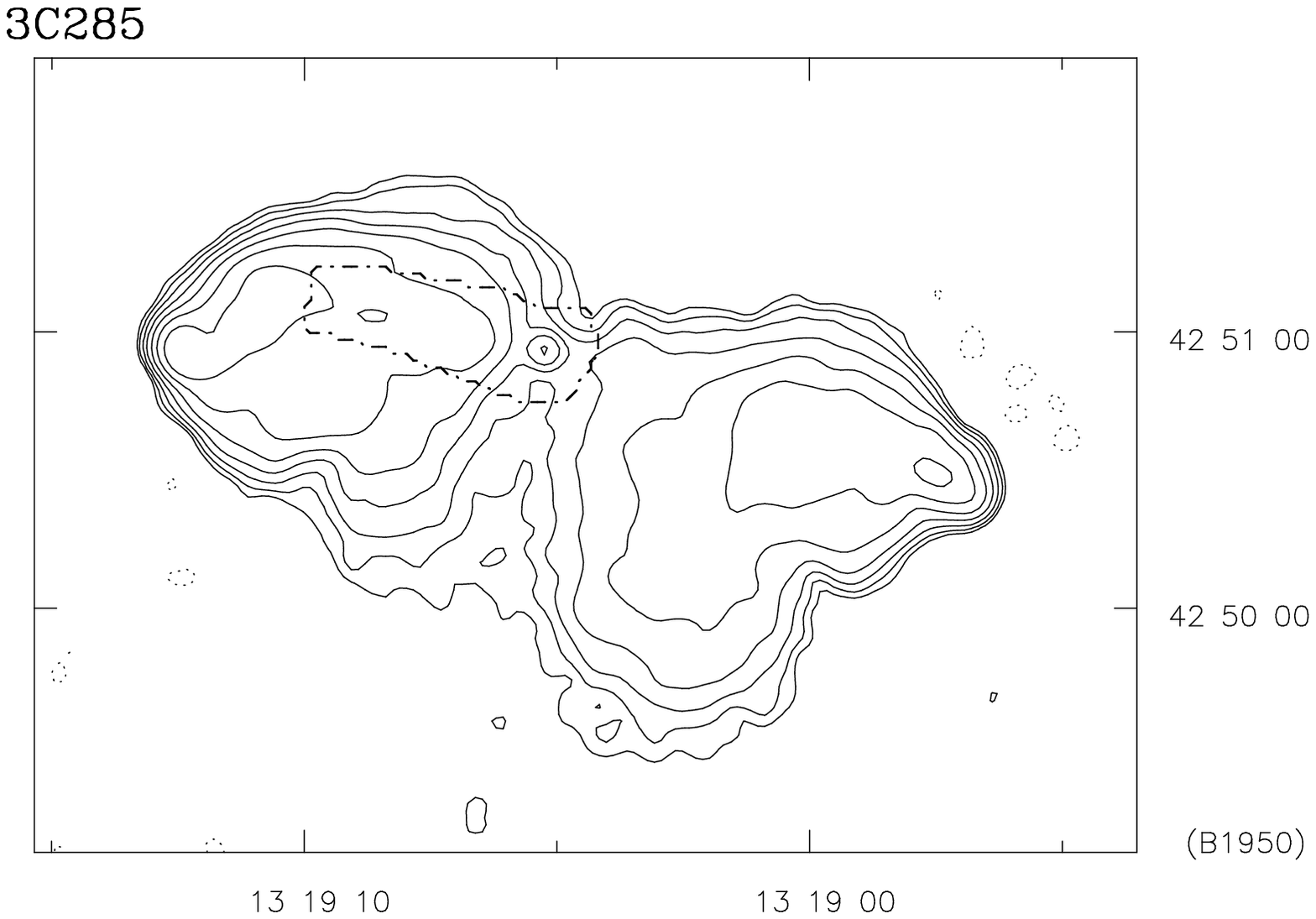,width=5.7cm, angle=0}} 
  \centerline{
    \epsfig{figure=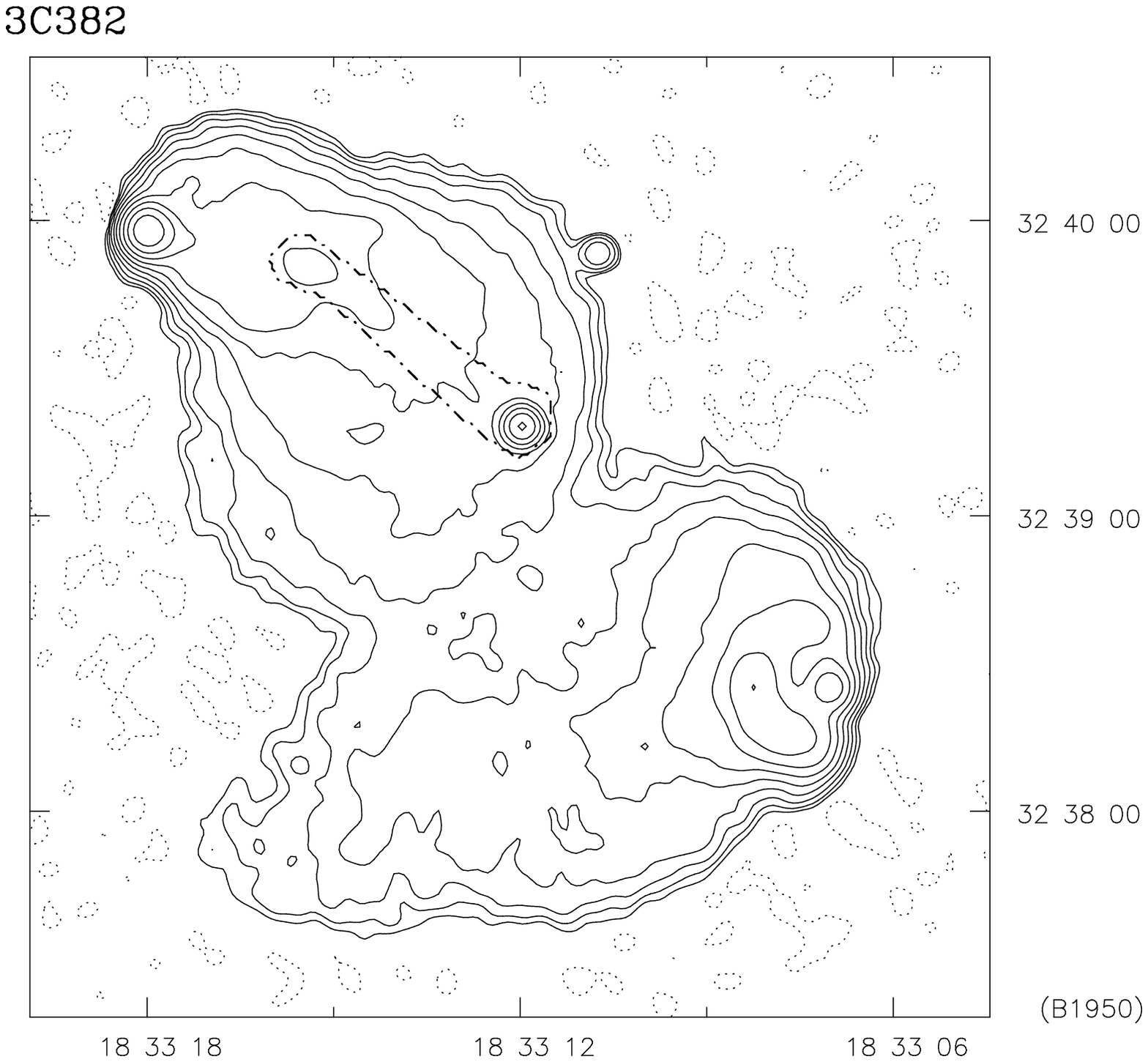,width=5.2cm, angle=0}
\raisebox{5.0cm}[0cm][0cm]{\epsfig{figure=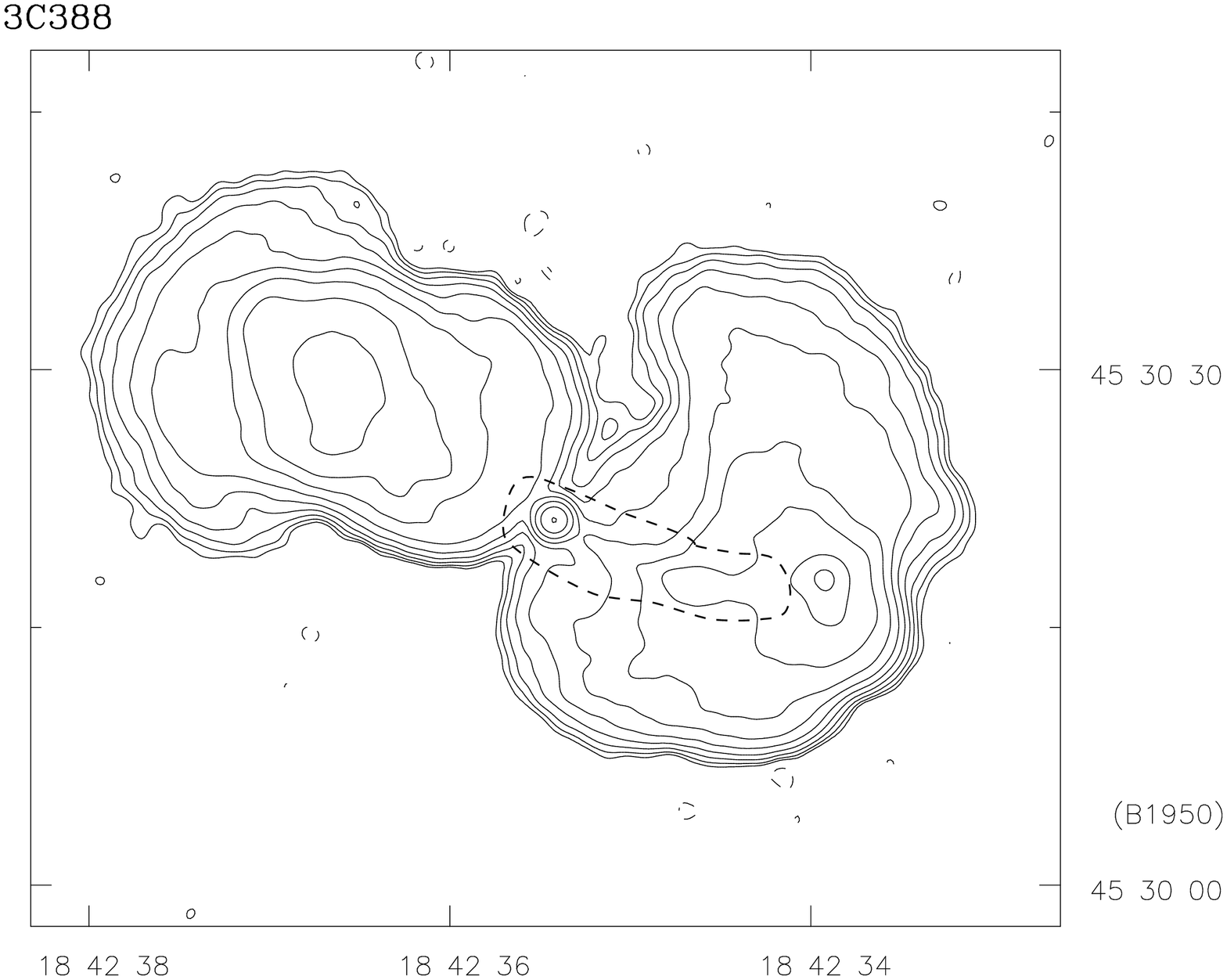,width=5.0cm,angle=-90}}} 
  \centerline{
    \epsfig{figure=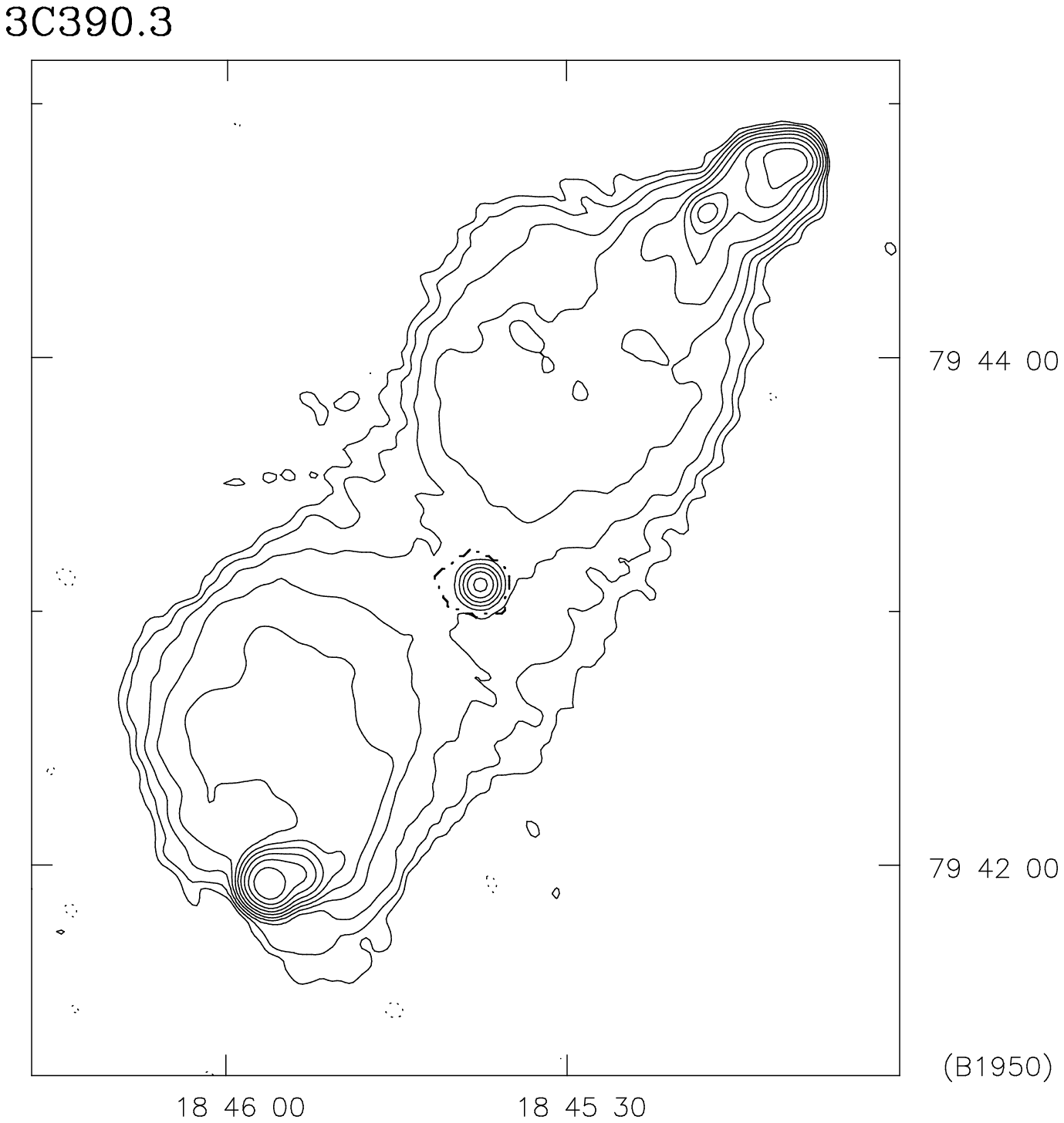,width=5.2cm, angle=0}
        \raisebox{5.0cm}[0cm][0cm]{\epsfig{figure=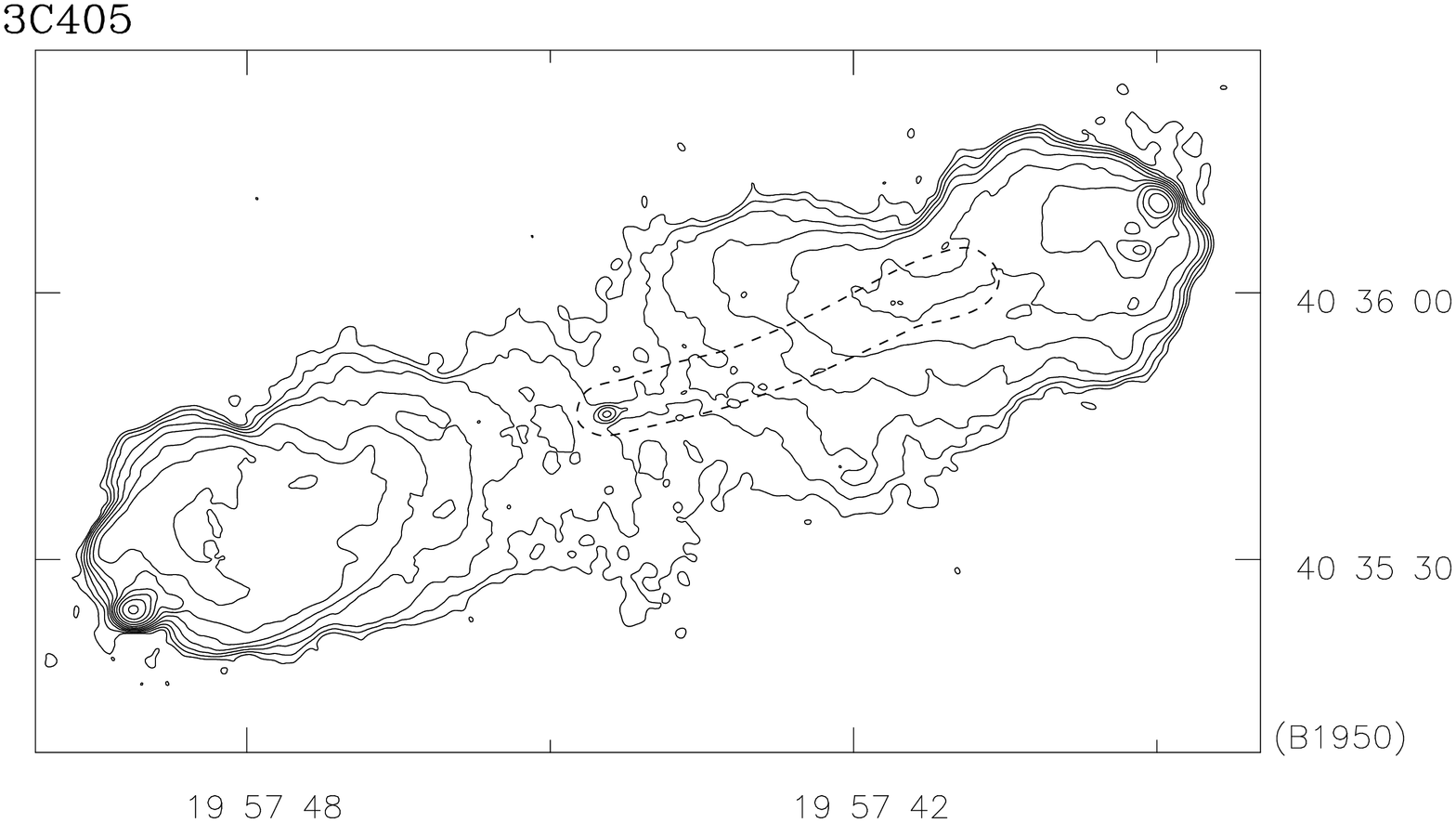,width=5.2cm, angle=-90}}}
  \centerline{
    \epsfig{figure=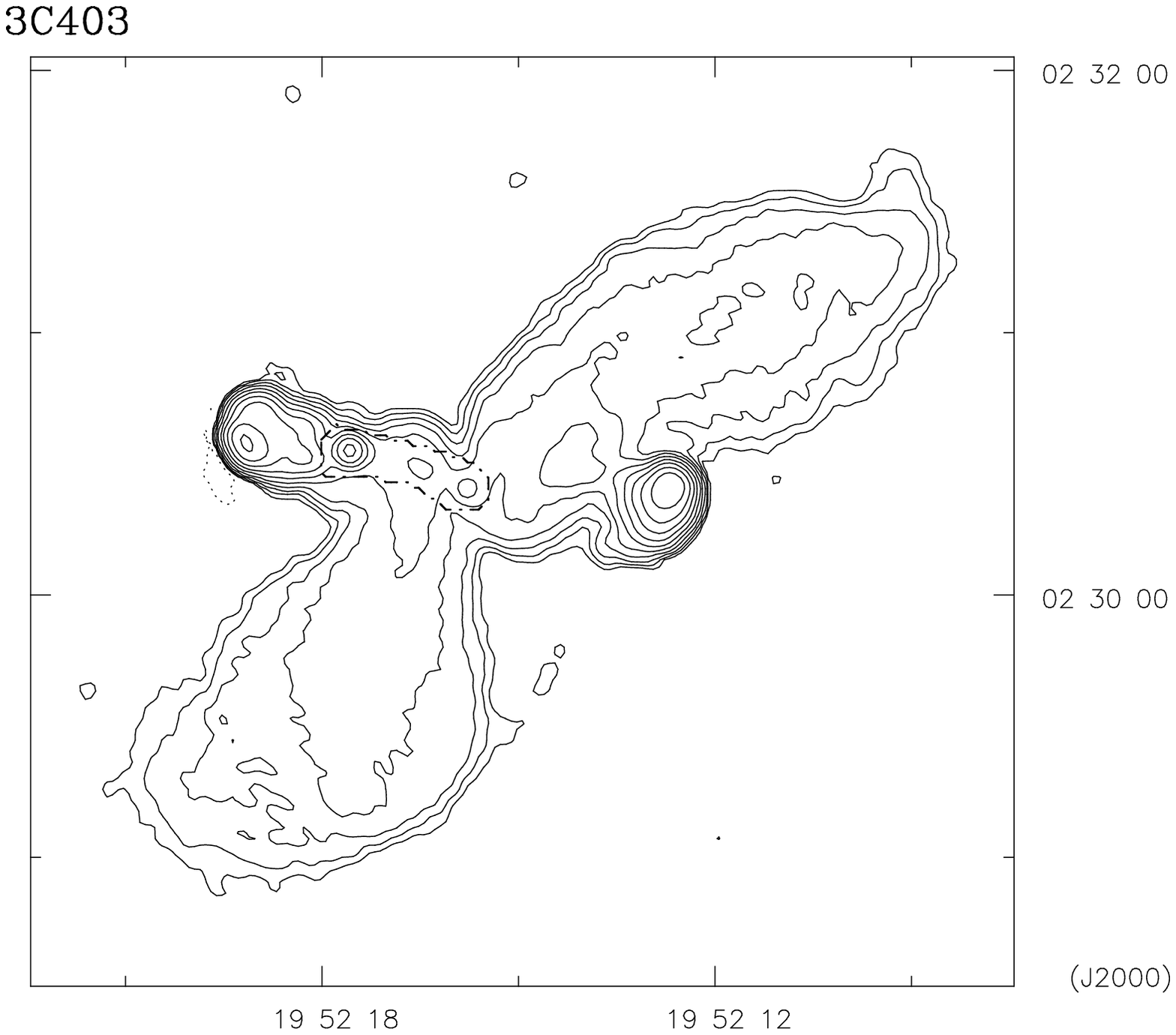,width=5.0cm, angle=0}
    \epsfig{figure=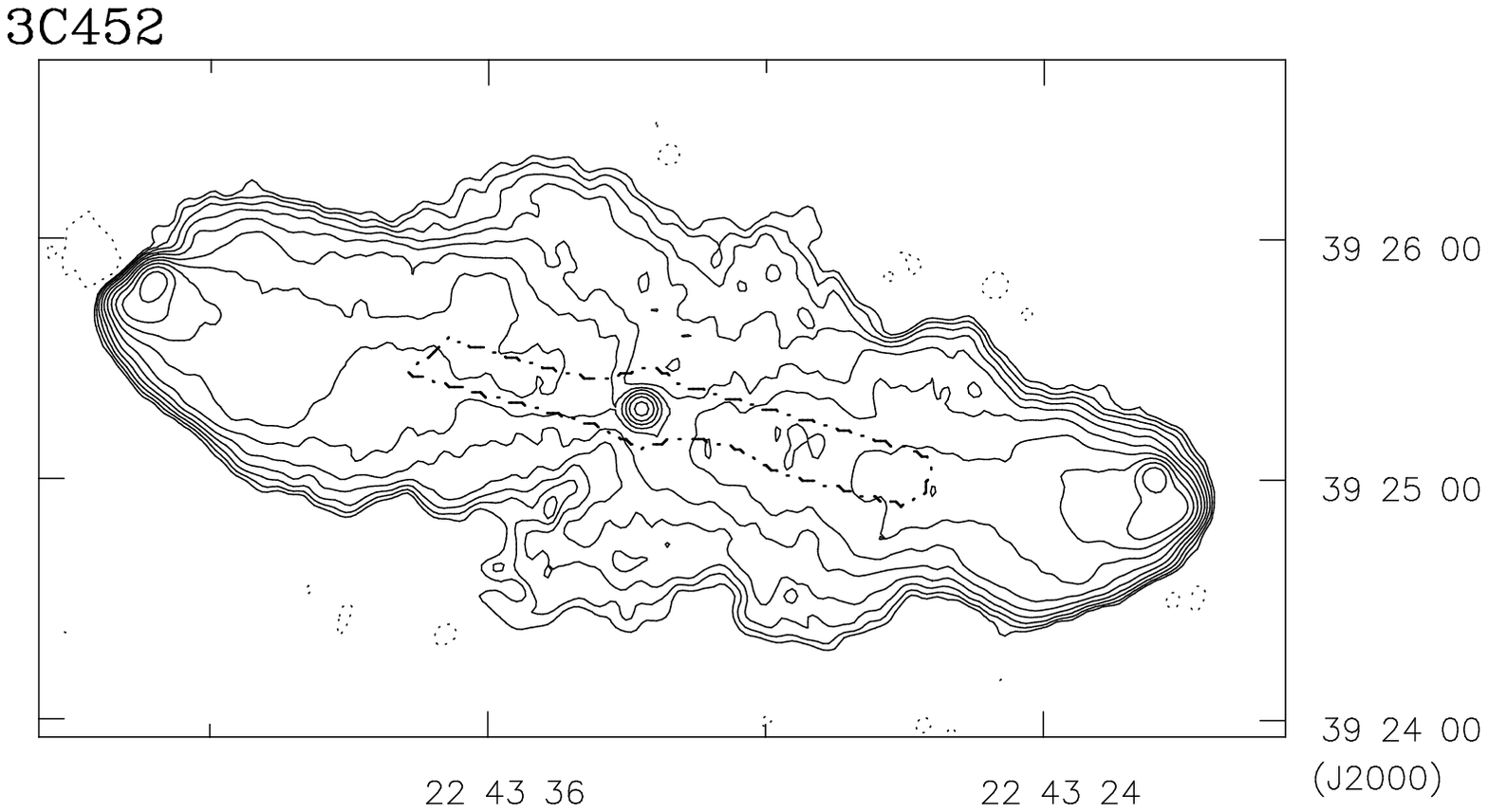,width=8cm, angle=0}} 
\caption{Images at 1.4~GHz of the sample. The contours are those used to
  divide the sources up into the surface brightness regions used in the
  spectral index calculations. The dotted lines indicate excluded areas.}
\label{fig:images}
\end{figure*}

The resulting 1.4~GHz images are shown in
Fig.~\ref{fig:images}. Table~\ref{tab:images} shows the beam sizes used,
the noise levels attained and the contouring factor for the displayed images.
The lowest contour is at 3$\sigma$ and the contours are logarithmically
spaced. For completeness all 1.4~GHz images are shown.

\begin{figure*}
  \centerline{
    \epsfig{figure=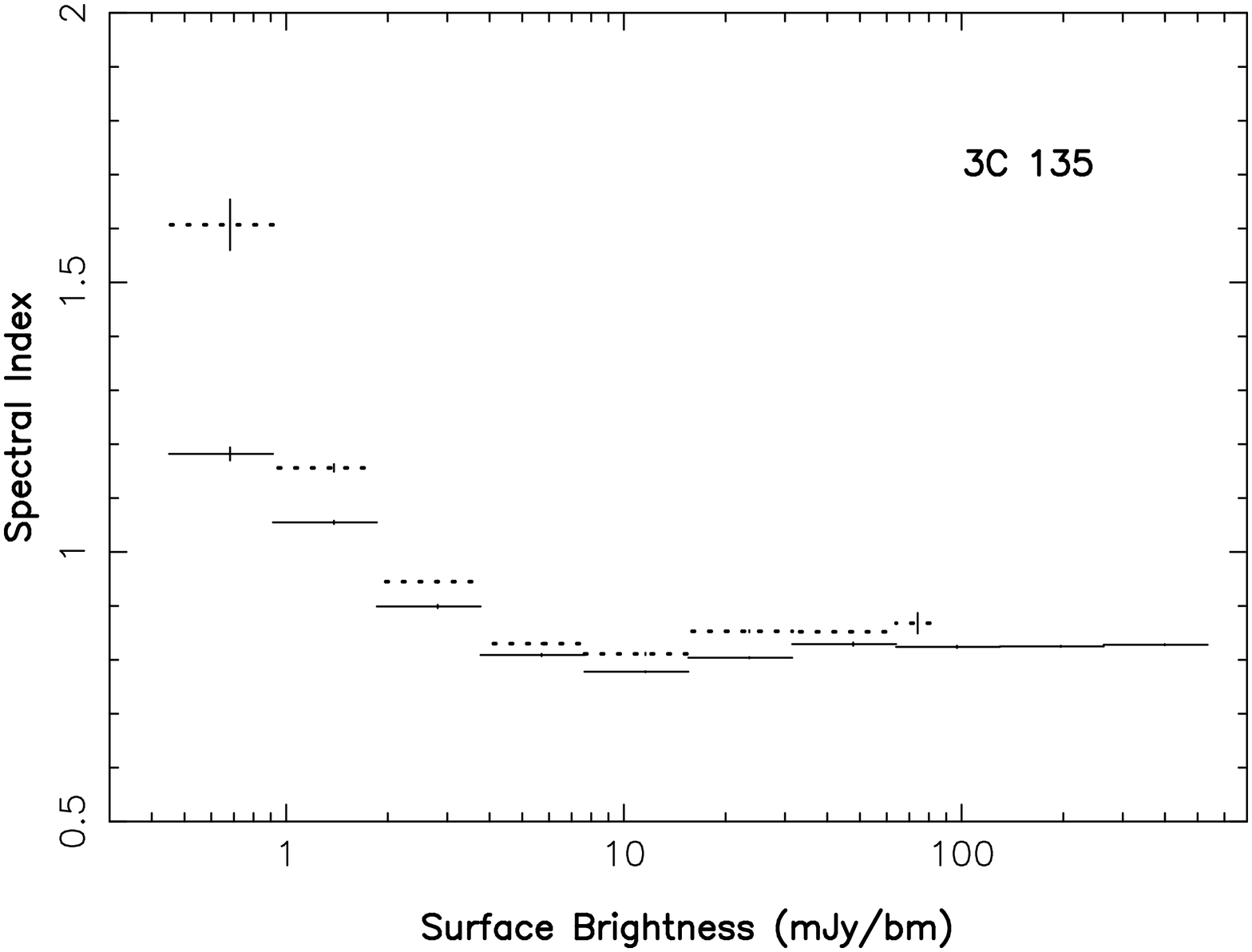,
      width=4.6cm, angle=-90}
    \epsfig{figure=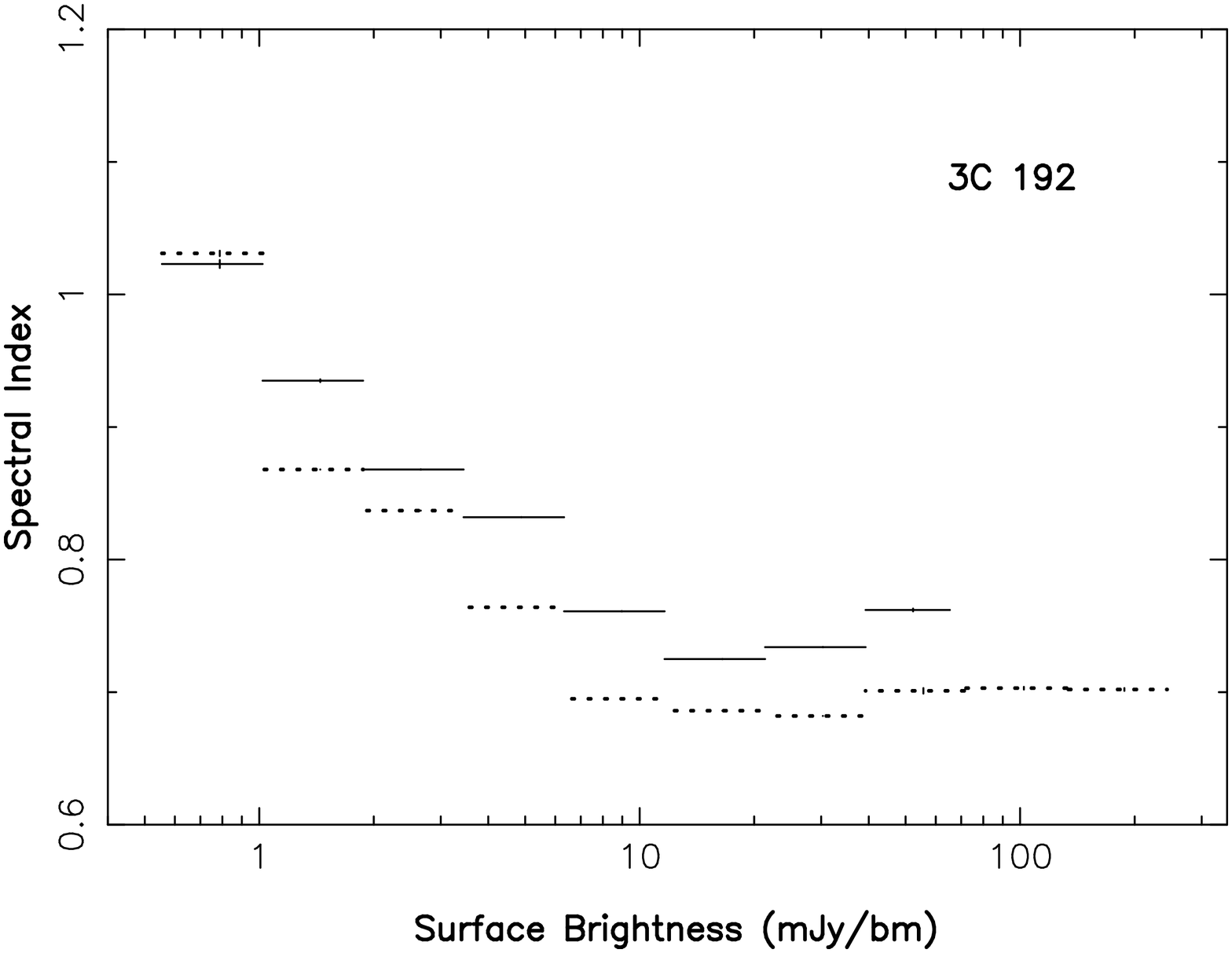,
      width=4.6cm, angle=-90}} 
  \centerline{
    \epsfig{figure=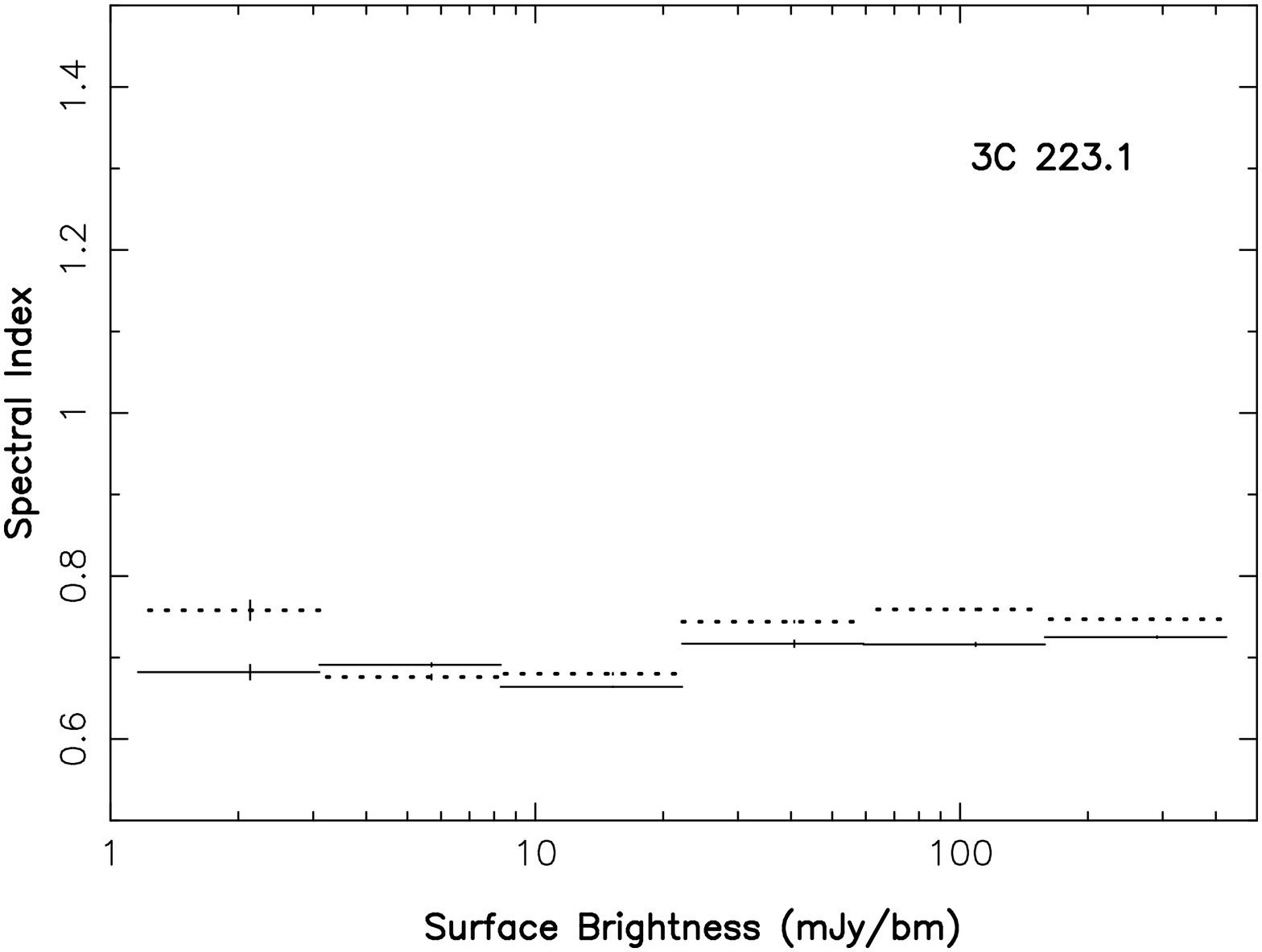,
      width=4.6cm, angle=-90}
    \epsfig{figure=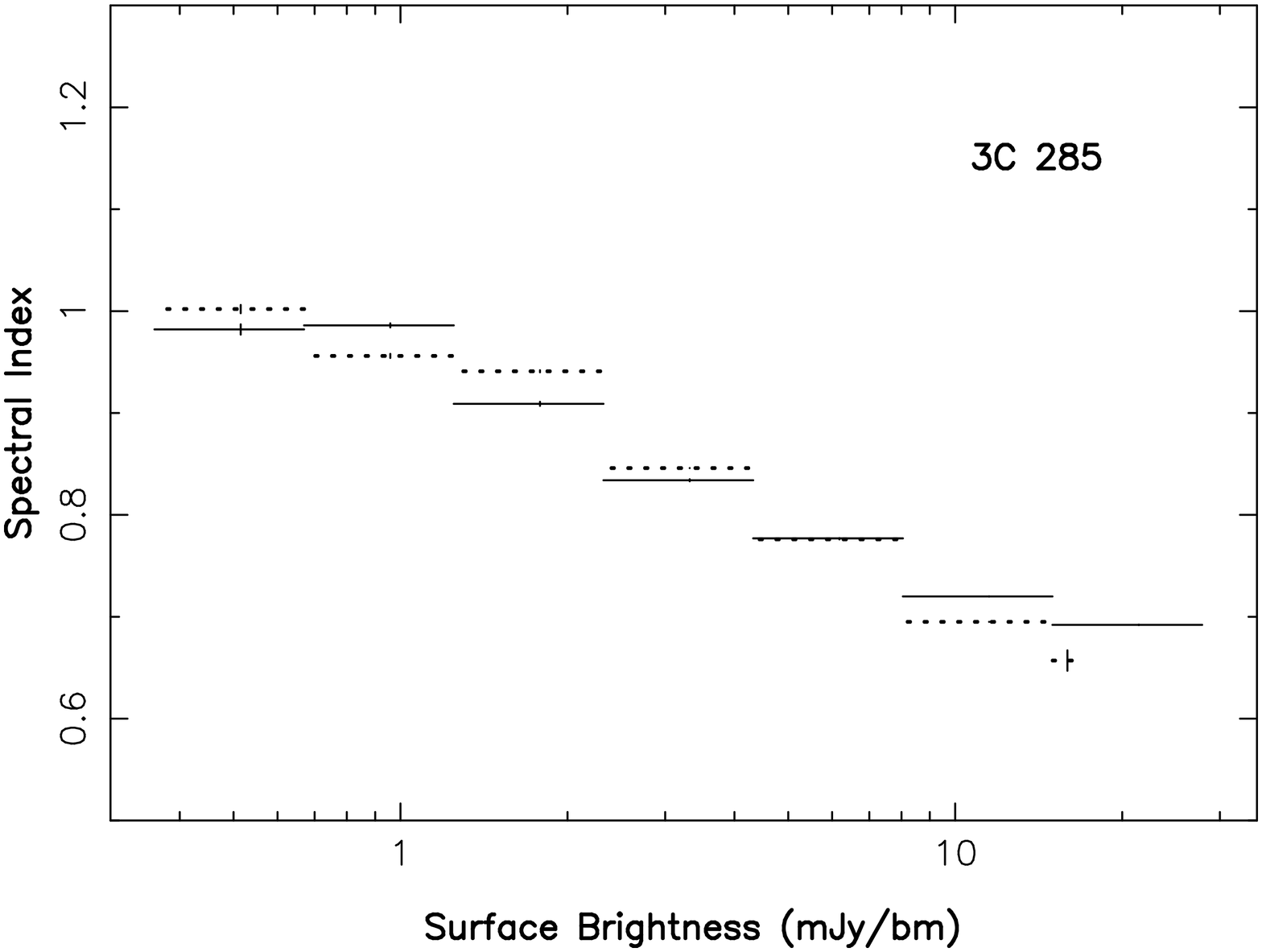,
      width=4.6cm, angle=-90}} 
  \centerline{
    \epsfig{figure=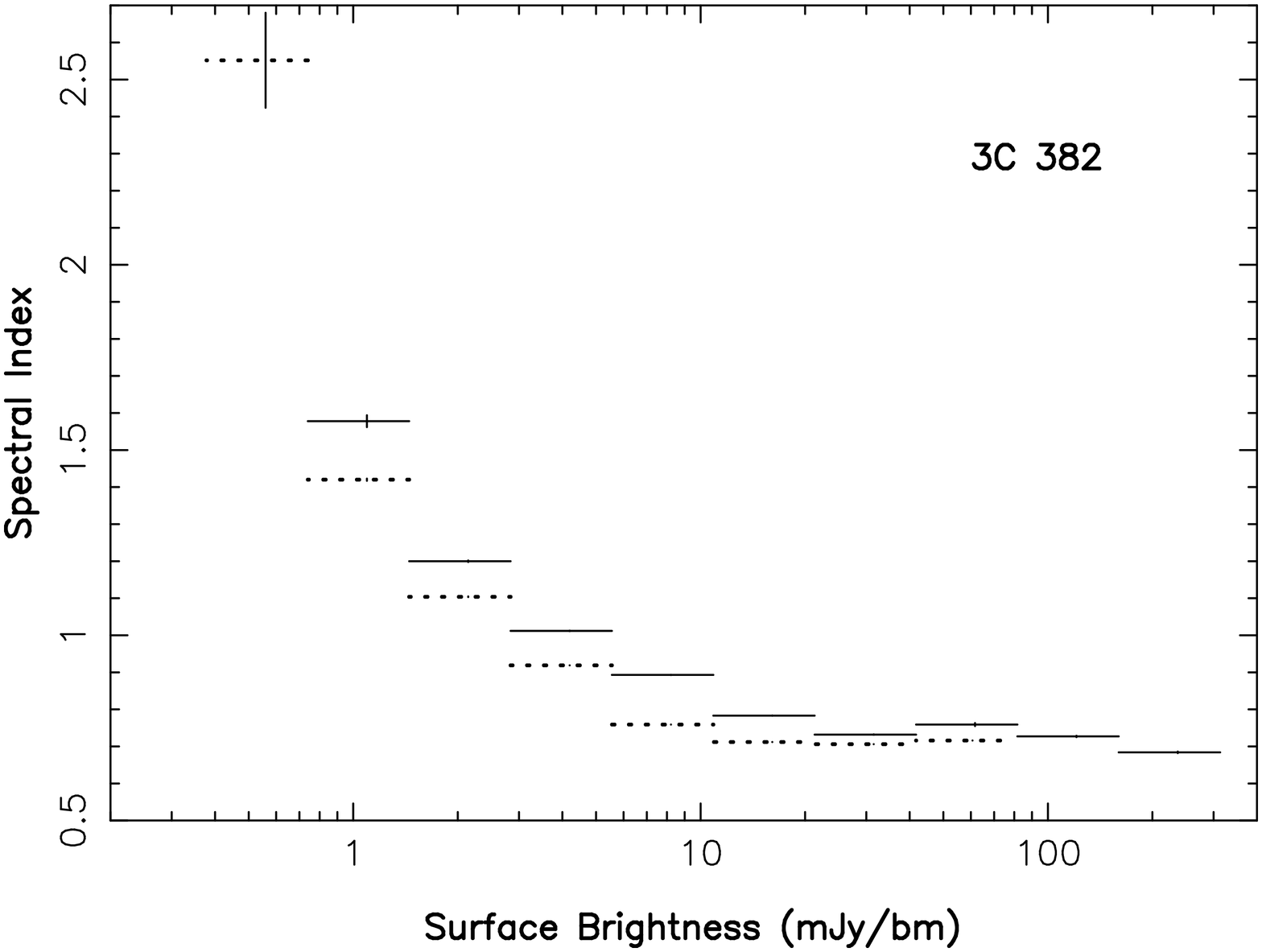,
      width=4.6cm, angle=-90}
    \epsfig{figure=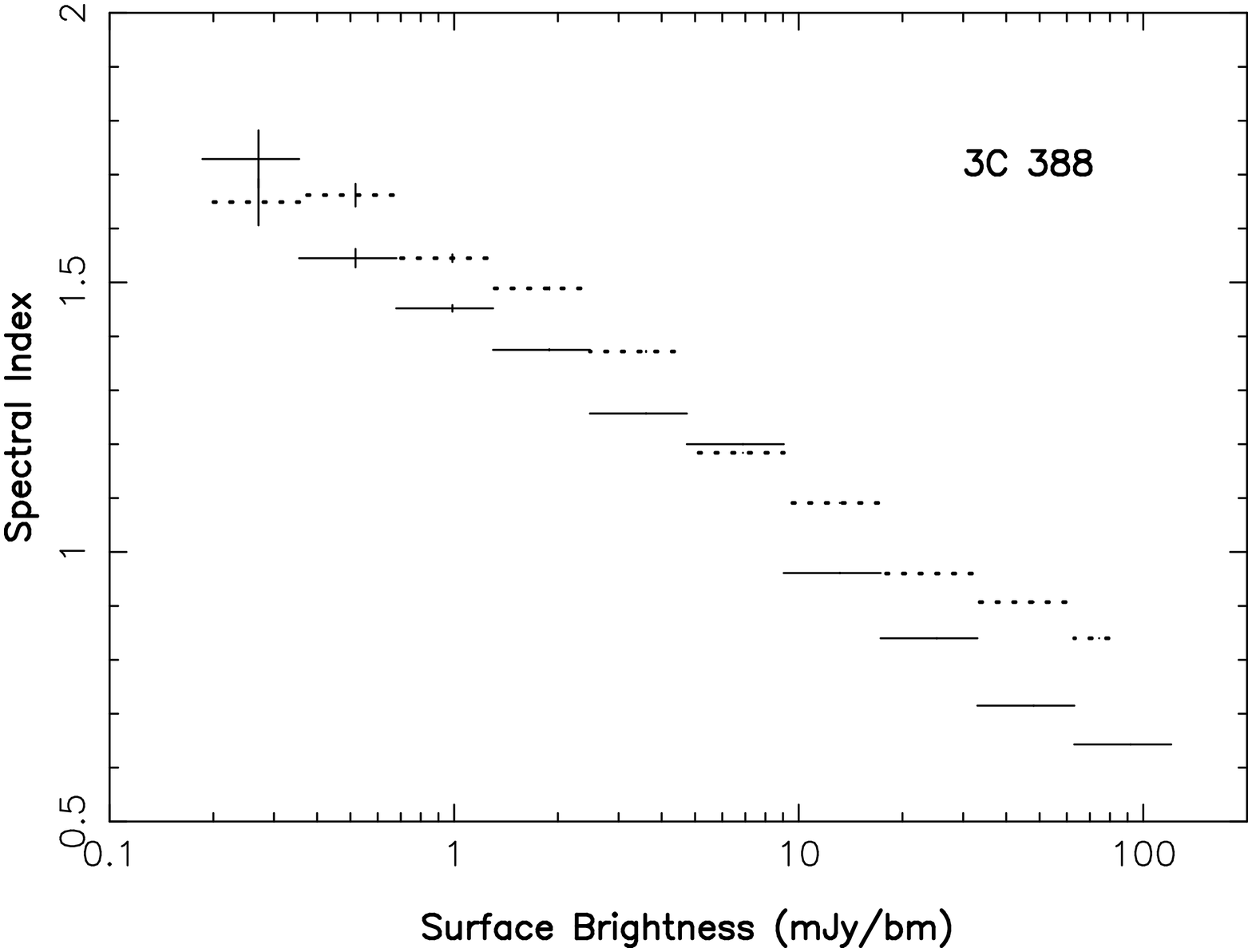,
      width=4.6cm, angle=-90}} 
  \centerline{
    \epsfig{figure=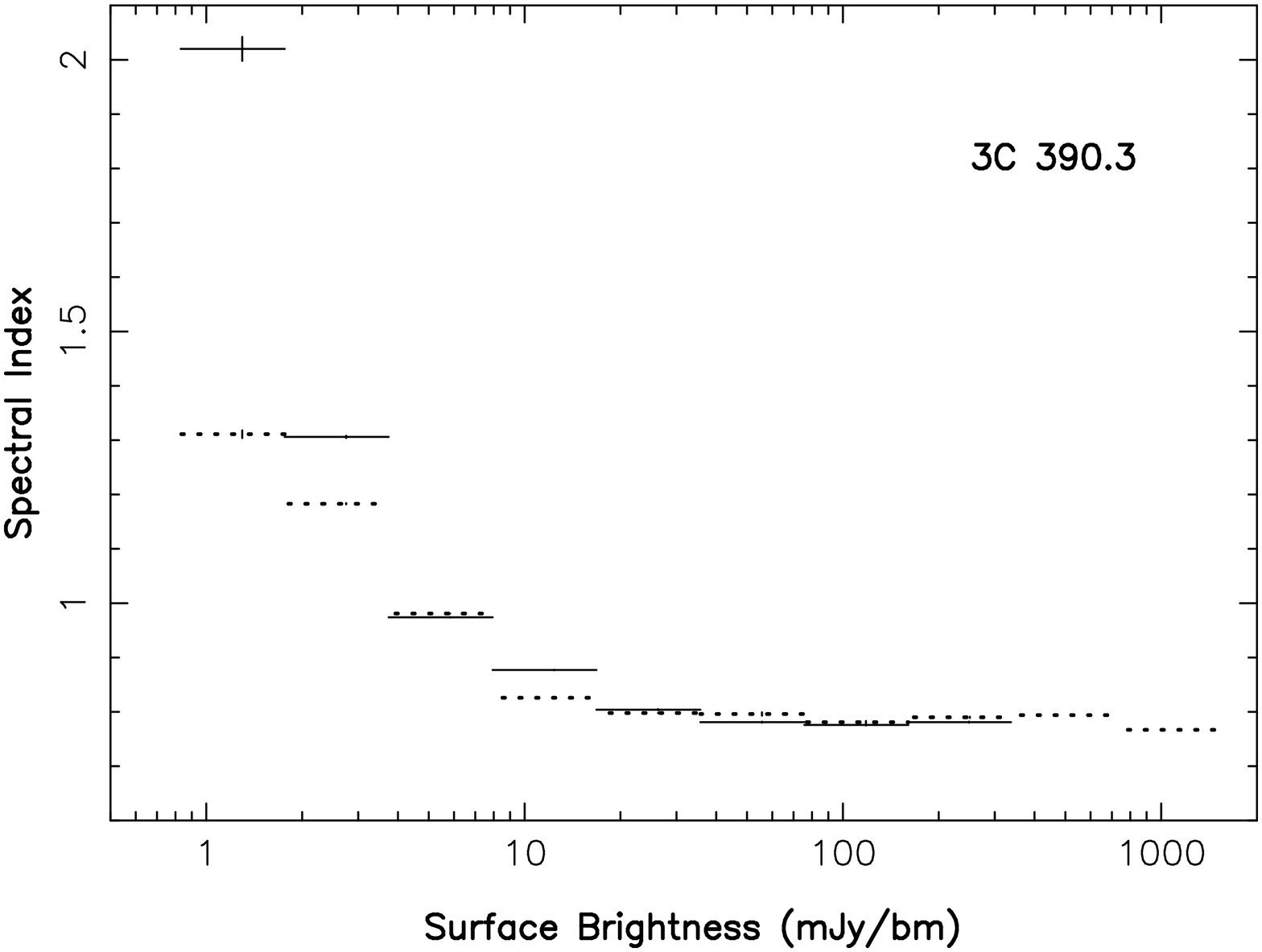,
      width=4.6cm, angle=-90}
    \epsfig{figure=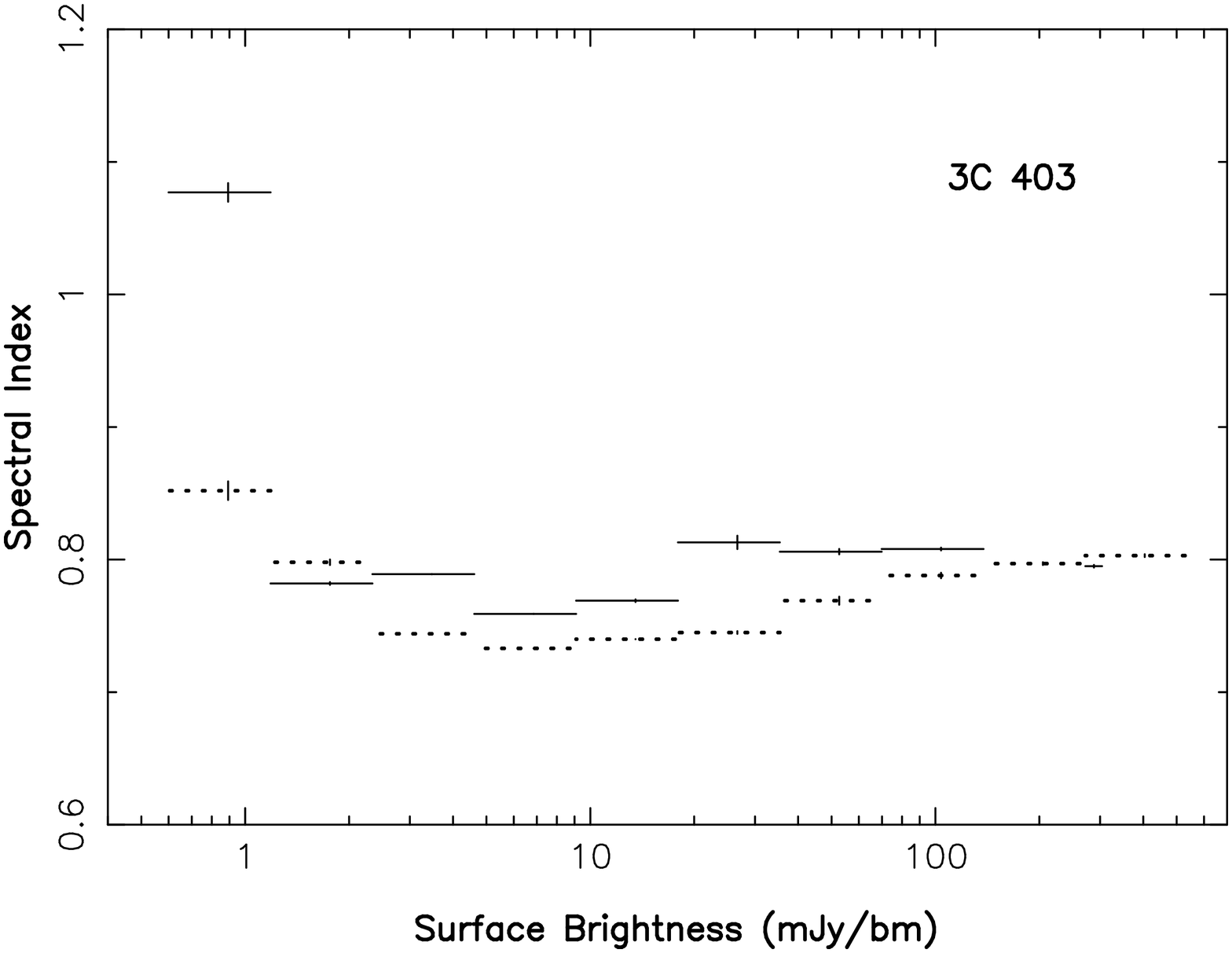,
      width=4.6cm, angle=-90}} 
  \centerline{
    \epsfig{figure=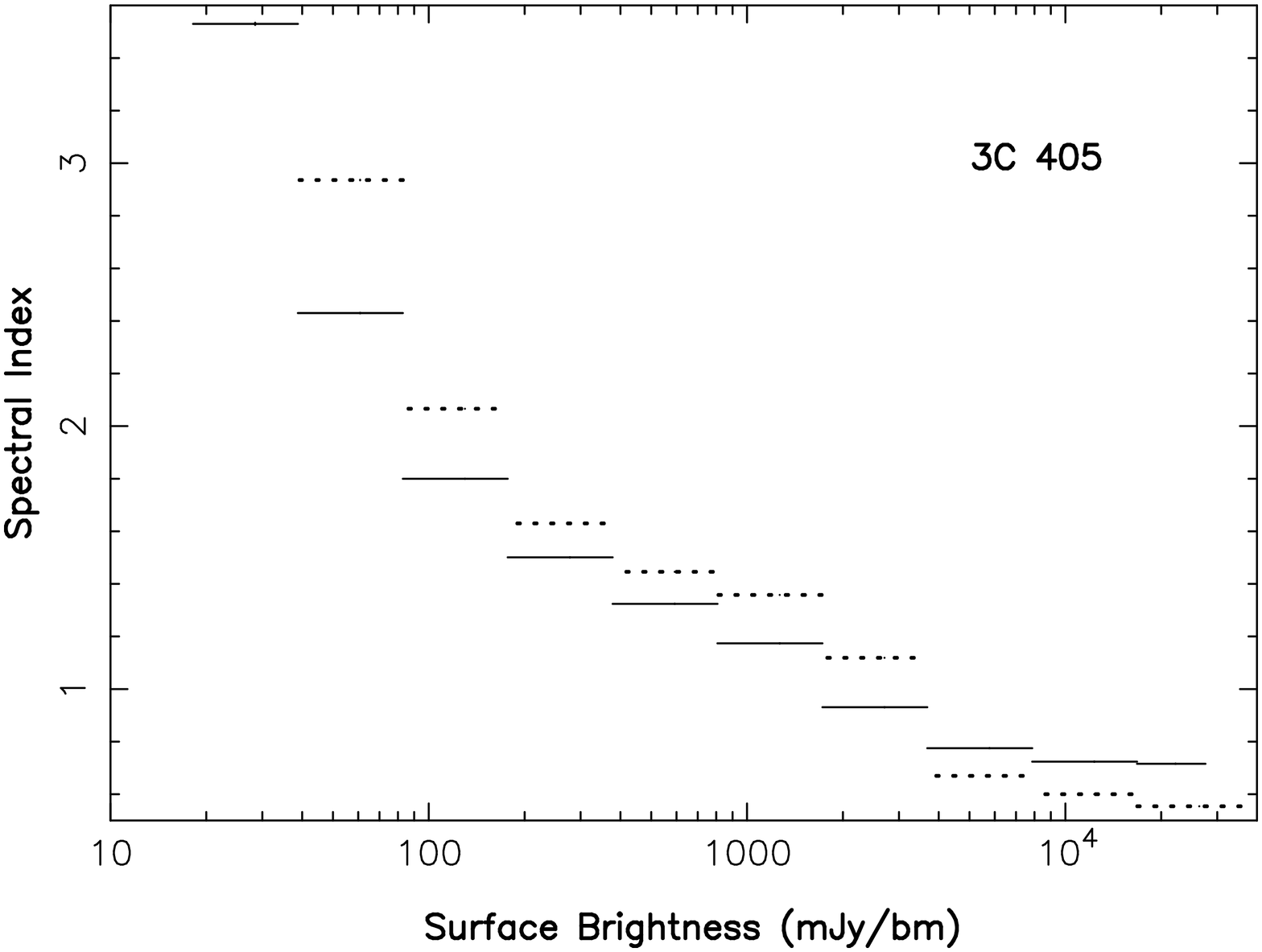,
      width=4.6cm, angle=-90}
    \epsfig{figure=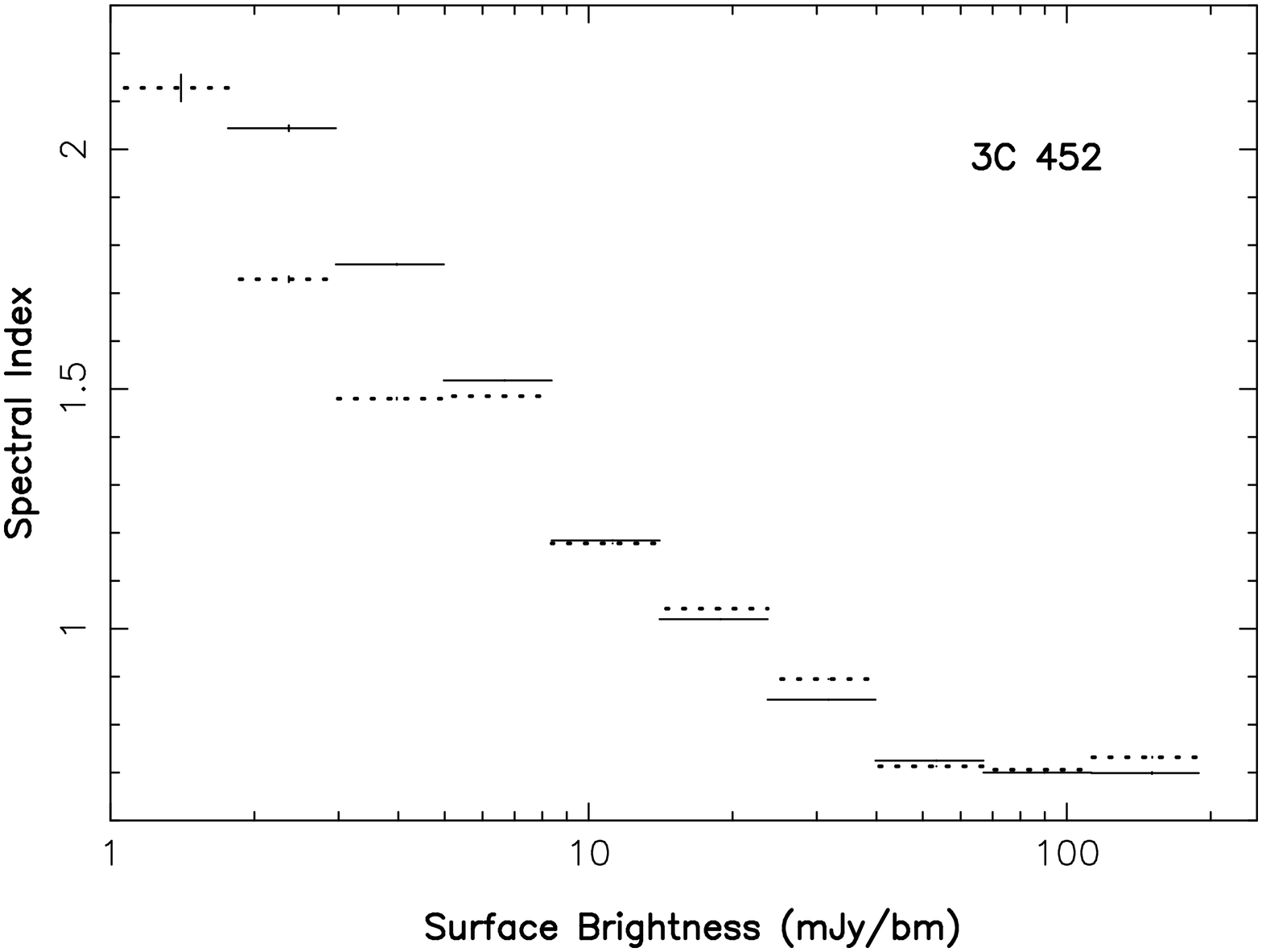,
      width=4.6cm, angle=-90}}
\caption{Plots of the two-point spectral indices versus surface
  brightness on the jet (solid lines) and counter-jet (dotted lines)
  sides of the sources. The frequencies used in the spectral index
  calculation are indicated.}
\label{fig:plots}
\end{figure*}

\section{Asymmetries}

\subsection{Analysis}

Asymmetry in spectral index was assessed by comparing regions 
of equal surface brightness on opposite sides of each source. The procedure is 
explained in more detail in Paper~1: in essence,  the image is cut along total 
intensity contours in the 1.4~GHz image (the contours being those shown in 
Fig.\ref{fig:images}) and these contours are used as templates for the image 
at the higher frequency. The spectral index is calculated for each region 
between two brightness levels (a `bin') from the ratio of flux densities in
that bin. 

In the radio galaxies, the lobes reach all the way to the middle of
the source.  The dividing line between the two lobes is usually
obvious, but in some sources (e.g. 3C192) it is not clear how the low
surface brightness bins should be divided between them.  All of these
sources have central troughs in their surface brightness
distributions, often coinciding with clear changes in the polarization
structure suggestive of the meeting of two volumes of back-flowing
plasma (Black \etal\ 1992; Leahy \etal\ 1997).  We have therefore
divided the sources along these intensity minima. In the two
``winged'' sources in the sample, the wings were associated with lobes
as follows: 3C223.1 NE (jet side) lobe with NW wing; 3C403 NE (jet
side) lobe with SE wing.

Note that the jets are not clearly visible in Fig.~\ref{fig:images}:
higher resolution images (Black \etal\ 1992; Leahy \etal\ 1997) were
used to identify regions with significant jet emission and these were
excluded from the spectral analysis, as in Paper~1.  The excluded
regions are indicated by dashed lines in Fig.~\ref{fig:images}.

The spectral indices for the various surface brightness bins are shown
in Fig.~\ref{fig:plots}. The horizontal lines represent the range of
surface brightness in which the spectral index was calculated. The
errors due to noise in the images are shown by vertical lines;
systematic errors in the flux calibration will affect both sides
equally and will therefore not change our conclusions.
Fig.~\ref{fig:plots} shows the usual steepening of the spectrum from
the hotspots to the low-brightness regions of the lobes, except in
the two winged sources (see below).  

\subsection{Correlations with jet sidedness}

There is no trend of spectral index with jet side at any brightness
level. Indeed, any differences between the spectra of the high
surface-brightness regions on the jet and counter-jet sides are not
merely random: in 7/10 sources (including the two BLRG 3C382 and
3C390.3) they are insignificantly small.  The exceptions are 3C192,
3C388 and 3C405, but even here the spectral-index differences are
$\leq$0.1.  In most cases the spectral index converges to a well-defined
value over the top two or three bins, showing that the `true' hotspot
dominates the flux of these regions even at this resolution. The
similarity of the hotspot spectrum on both sides of the source gives
evidence for a single injection spectrum in each source.

\subsection{Correlations with lobe volume}

For the quasar sample of Paper~1 we found that the longer lobe had the
flatter spectrum in 8 sources out of 9. Here ``length'' is defined as
the distance from the core to the furthest part of the $3\sigma$
contour and the spectrum is evaluated for the lowest three surface
brightness bins.  The most natural explanation for this correlation is
that synchrotron losses have proceeded more slowly in the larger lobe
because the magnetic energy has spread into a larger volume, so that
there is a smaller average magnetic field.  We looked for a similar
effect in the radio galaxies, but found no correlation of the spectral
indices for the 3 lowest surface brightness bins with either: (a) the
lobe length, defined as above, or (b) the area of the lobe enclosed
within the 3$\sigma$ contour of the 1.4~GHz image.

It is perhaps not surprising that a correlation is more difficult to
detect in radio galaxies: the quantity which seems physically most
relevant is the lobe volume and, as the galaxies are more symmetrical
(Fig.~\ref{fig:images}; also Best \etal\ 1995), more care is required is
required in its estimation.  There is no entirely satisfactory method
for estimating the volume. For a start, one must assume that the lobe
is as thick as it is wide in the image plane. Furthermore, it is hard
to find a prescription for the width of a lobe which is at the same
time reasonable and objective. For example, it is evident from
Fig.~\ref{fig:images} that the width measured from the line joining
the hotspots to the $3\sigma$ contours is a reasonable measure in
straight sources such as 3C452 or 3C390.3, but makes no sense at all
for 3C135, 3C192 or 3C403. Measures such as area$^2$/length are
unsatisfactory for similar reasons.

We addressed the problem of bent sources using the following
method. The images at 1.4~GHz were first divided into the two lobes,
as in the spectral analysis. A curve was drawn starting from the
hot-spot along the ridge of the radio emission in each lobe. This
curve was approximated by a number ($\approx$10) of line segments of
equal length $l$.  Perpendiculars were then drawn from each line
segment to the $3\sigma$ contours on either side of the lobe and the
average of the two lengths was taken to be the radius of the lobe,
$r_i$.  The lobe was then approximated as a set of cylinders of height
$l$, so the volume was taken to be $\pi l \Sigma r_i^2$. From this the
fractional volume asymmetry was readily calculated. In the case of
3C403 the W lobe and the NW wing were distinguished from each other
for the purposes of the measurement of the nominal `diameters' (in
both winged sources, the line used as the ridge followed from the
hot-spot to the tip of the wings in a smooth curve).

Further problems occur in the quasar sample as the lobe emission
disappears into the noise some distance from the core.  The difference
in redshift between the two samples means that the emitted frequency is
significantly higher for the quasars, and so that the lobes will be more
difficult to detect for a given amount of spectral curvature in the rest
frame (see below). For four of the quasars, Leahy, Muxlow \& Stephens
(1989) presented 151MHz MERLIN images. These show emission extending
further towards the core, as do 1.4-GHz images at lower resolution. We
therefore believe that the lobe volume will be significantly
underestimated if the boundary of the observed emission in Fig. 1 of
Paper 1 is used.  In many quasars a rapid and substantial decrease in
apparent width occurs, at which point the source no longer has a sharply
defined edge. This is most likely to reflect the fading of the
lobe into the noise, rather than the real source structure.  In a few cases,
there is no detectable lobe flux density in the regions closest to the
core.  In these situations, we used the last diameter measured before
the sudden decrease in apparent width as the diameter all the way back
to the core.  

The extrapolation adopted attempts, as far as is possible, to take
account of the dependence of observed source structure on spectral index
(flatter spectrum lobes will not fade into the noise so rapidly, and may
therefore have a larger region detected above the noise).  As a test of the
reliability of the method, we compared the volumes with values
calculated directly from 151~MHz images where these were available, and
found that the calculated asymmetry was in the same sense in all cases.
Differences in independent estimates of $V_j/V_{cj}$ were $<$30\%. In
the case of the radio galaxies there are few regions where the sources
appear to fade into the noise: a sharp edge is generally observed,
indicating that we are likely to be observing the true boundary of the
source.

\begin{figure}
  \centerline{
    \epsfig{file=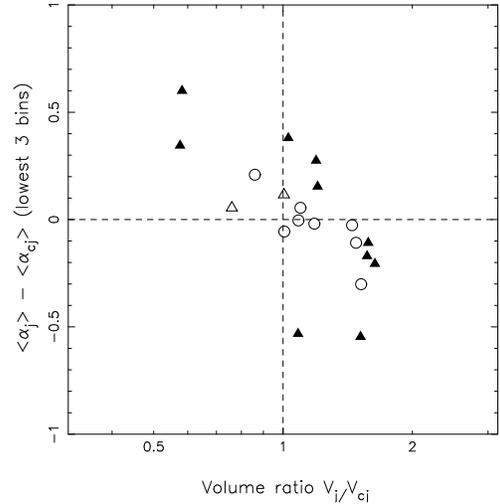
      , height=7cm, angle= -90, clip=}}
\caption{Spectral index differences of lowest three bins for all
sources against `volume ratio'. Larger lobes with flatter spectra
lie in the second and fourth quadrants. Quasars (filled triangles) and radio
galaxies (open symbols). BLRG indicated by open triangles.}
\label{fig:low-sb-vol}
\end{figure}

The result is shown in Fig.~\ref{fig:low-sb-vol}. In 7 out of 10
quasars, and 8 out of 10 radio galaxies, the larger lobe has the flatter
spectrum.  The Spearman Rank Test is an efficient method for testing the
significance of the effect, and it shows that the correlation is
significant at the 97\% level for the quasars and 99\% for the radio
galaxies independently ($r_s = 0.68$ \& $0.75$ respectively), and at
much greater than the 99\% level for the samples combined ($r_s =
0.71$). The radio galaxies are more symmetrical in both spectral index
and lobe volume than the quasars, as they are in lobe length. Although
there are obvious difficulties in the estimation of lobe volume, we
emphasise that the sources for which the calculated volume is most
questionable (the quasars) are precisely those which exhibit the secure
correlation between lobe length and spectral index (Paper 1).  Thus, there is clear
evidence that the spectral index of low surface brightness emission is
flatter in the larger lobe in both classes of source. \nocite{bes95}

\section{Differences between the radio galaxy and quasar samples}

Many comparisons have been made between radio galaxies and quasars, some
on much larger samples than ours, but not with spectral information at
high angular resolution, allowing a clear separation between hotspots
and regions of low surface brightness.

\subsection{Hotspot regions}

\begin{figure}
\epsfig{file=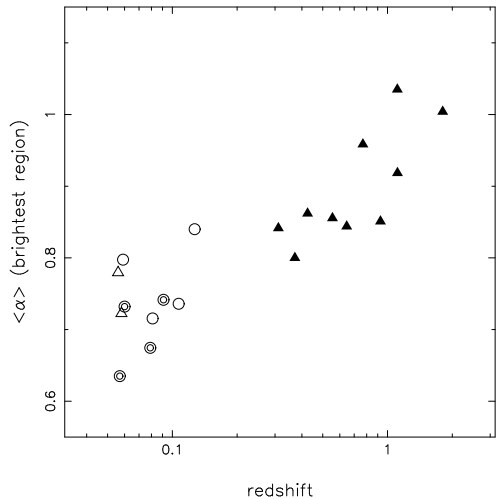, width
= 7cm, angle = -90}
\caption{Observed spectra of hotspots (brightest regions defined in text)
  averaged over jet and counter-jet sides. Quasars (filled triangles),
  narrow-line radio galaxies (circles -- those with inner circle
  indicate observations between 1.4 \& 5~GHz) and broad-line radio
  galaxies (open triangles). }
\label{fig:k-corr-hotspots}
\end{figure}

The correlation between spectral index and redshift or luminosity is
well established for the integrated emission of sources in the 3CR
sample.  For example, Laing \& Peacock (1980) showed that there is a
strong correlation of the spectral indices $\alpha_{750}^{1400}$ of the
integrated (non-core) emission of FRII galaxies with $P_{1400}$ and $z$.
As Jenkins \& McEllin (1977) had already shown that the fractional flux
in compact components ($<$15kpc) increased with increasing radio power,
Laing \& Peacock were able to infer that the correlation of spectral
index with radio power or redshift must be even stronger for the
hotspots, as they have flatter spectra than the extended lobes. This
effect is shown directly by our samples.  Fig.~\ref{fig:k-corr-hotspots}
confirms that the hot-spot spectra of the quasars are markedly steeper
than those of the radio galaxies. In this figure (and ensuing
discussion) the brightest region of both sides is taken as that brighter
than the lower bound of the brightest bin of the lobe with the lower
peak surface brightness. This is the procedure adopted in Paper 1 to
allow paired spectra in lobes with differing peak surface brightnesses.
Our discussion and conclusion below remain unaffected if we were to use,
instead, peak flux densities or highest surface brightness bins on both
sides.

\nocite{lai80b}\nocite{jen77}

\begin{figure}
\epsfig{file=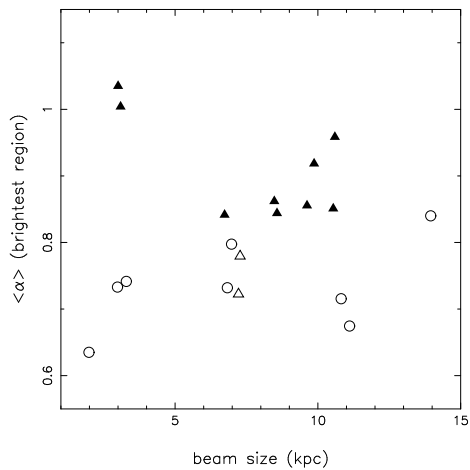, width
= 7cm, angle = -90, clip=}
\caption{Observed spectra of hotspots
  averaged over jet and counter-jet sides as a function of linear
  dimension of observing beam. A positive correlation would be expected
  if steeper spectrum `hotspots' were due to the inclusion of lobe
  material in hotspot region.   Symbols as before.}
\label{fig:beam}
\end{figure}

\begin{figure}
\epsfig{file=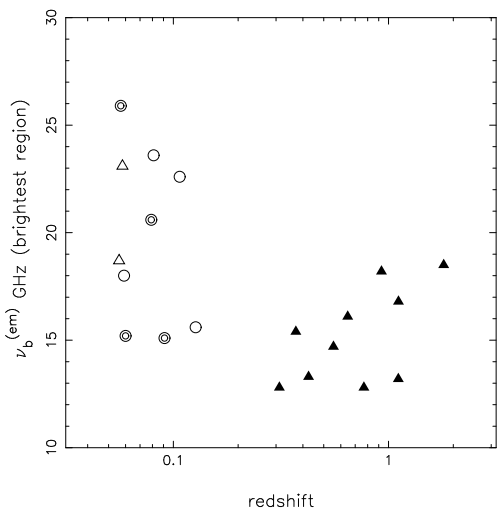, width
= 7cm, angle = -90, clip=}
\caption{
  Hotspot break frequencies (galaxy frame) calculated for all sources, using
  observed 2-point spectra and theoretical spectrum with sharp
  high-energy cut-off and injection index $\alpha_{inj}=0.5$ . The errors on $\nu_b$ at z~=~0 assuming an error on
the spectral index of $\sigma_\alpha$ = 0.3 are $\approx~\pm$~4~GHz for
$\nu_b$~=~20~GHz and $\approx~\pm$~0.5~GHz for $\nu_b$~=~7~GHz. The
errors on $\nu_b$ decrease with redshift, due to the spectral
curvature. Symbols as before.}
\label{fig:break-hot}
\end{figure}

Our observations of the radio galaxies are at lower angular resolution
than the observations of the quasars. As the decrease in resolution
approximately matches the difference in mean redshift between the
samples, differing linear resolution cannot account for the hotspot
spectral differences found between the samples. (This might occur if the
higher redshift objects were observed with a beam of larger linear
dimension, resulting in inclusion of more (steep spectrum) lobe material
in the hotspot region.) Fig.~\ref{fig:beam} shows the distribution of
linear beam sizes used. There was also no dependence of hotspot spectra on
the beam size as a fraction of the overall source size.

In sources whose spectra steepen at higher frequencies as a result of
synchrotron losses, a correlation between observed $\alpha$ and
redshift will necessarily exist, since we observe the high-$z$
sources at higher emitted frequency. It has been less clear whether
there is an additional contribution affecting the rest-frame spectral
index of the integrated emission and, if so, whether the fundamental
variable is redshift or luminosity.

While we do not doubt that simple redshifting of a curved spectrum
contributes to the observed $\alpha$ -- $z$ relation for hotspots
(Fig.~\ref{fig:k-corr-hotspots}), we believe that it cannot account
for the whole effect.  Observations at several frequencies covering a
wide range could resolve this issue directly but in their absence we
must assume a functional form for the rest-frame spectrum and
calculate the resulting two-point spectral indices as functions of
observing frequency and wavelength.  The most obvious functional form
is the theoretical spectrum of a power-law electron energy
distribution which has suffered synchrotron losses (cf. Myers \&
Spangler 1985\nocite{mye85}; Leahy \etal\ 1989).  We assumed that
electrons are injected with with an isotropic momentum distribution
and a power-law energy spectrum corresponding to $\alpha=0.5$ up to a
sharp cutoff and that they undergo pitch-angle scattering (cf. Jaffe
\& Perola 1973).  The resulting spectrum in the emitted frame is
parameterized by a single break frequency $\nu_b$.  Values of $\nu_b$
were then computed from the mean spectral indices of the brightest
bins of each lobe.

There is a clear difference between the radio galaxies and the quasars,
in the sense that hotspots in radio galaxies show higher break
frequencies (less synchrotron loss).  Other assumptions about the shape
of the rest-frame spectrum lead to qualitatively similar results. If we
assume that there is no pitch angle scattering (Kardashev 1962;
Pacholczyk 1970), the diagram looks similar but with 7~GHz $<\nu_b <$
15~GHz. The assumptions of a steeper injection spectrum leads to higher
$\nu_b$, particularly for objects with straight spectra -- thus
accentuating the difference between radio galaxies and quasars.
Similarly, the intrinsic difference between the classes would also be
increased, if we assumed a more realistic model in which the
relativistic electrons are injected continuously into the hotspots, as the
predicted curvature of the spectrum will be more gradual and the effect
of redshift diminished. \nocite{pac70,kar62,jaf73}

We therefore conclude that there is a difference between the rest-frame spectra
of the hotspots in the two samples. The simplest explanation is that the more
powerful hotspots have stronger magnetic fields and therefore faster
synchrotron loss. 

\subsection{Extended regions}

\begin{figure}
(a)\epsfig{file=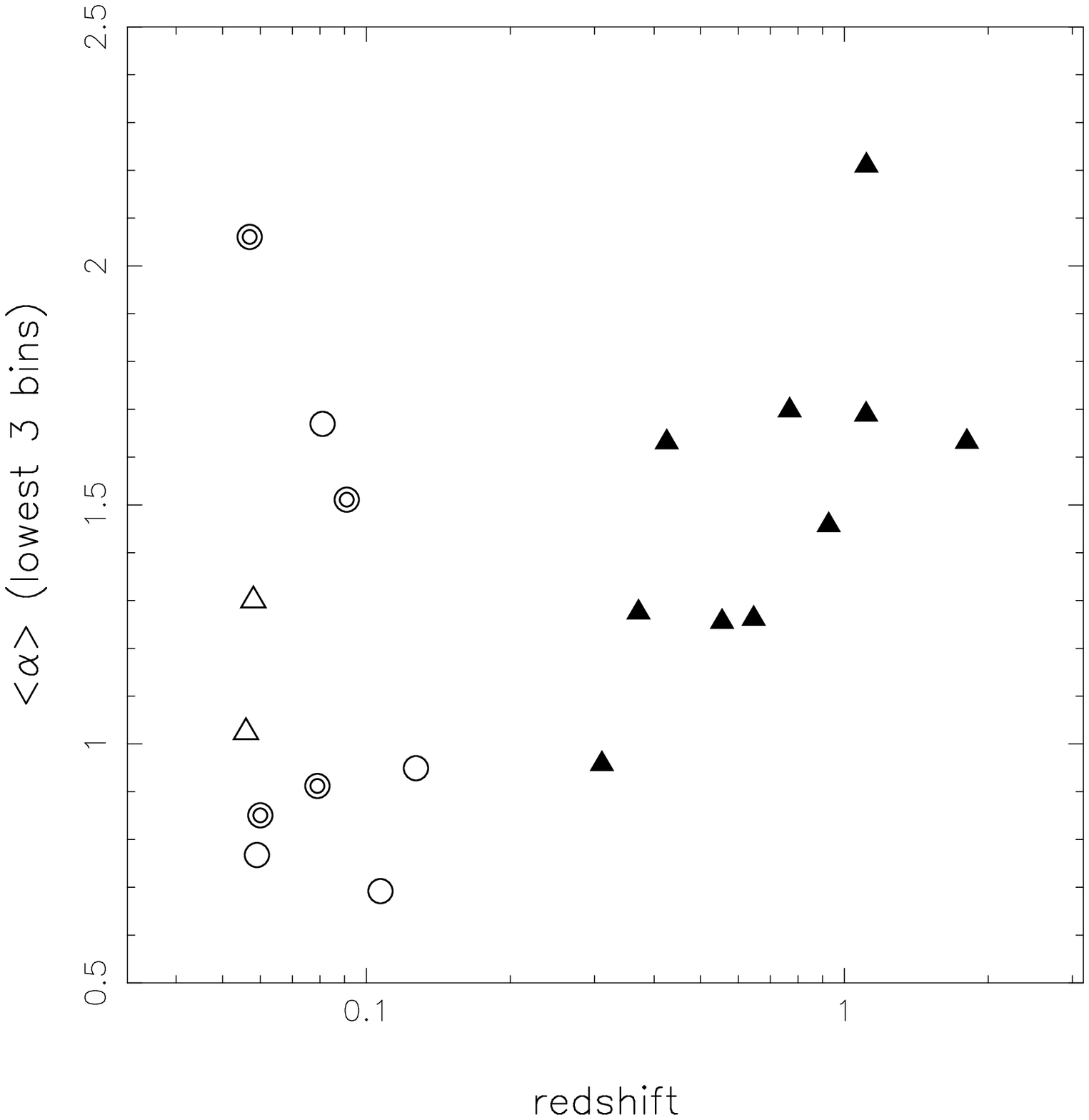, width
= 7cm, angle = -90}
(b)\epsfig{file=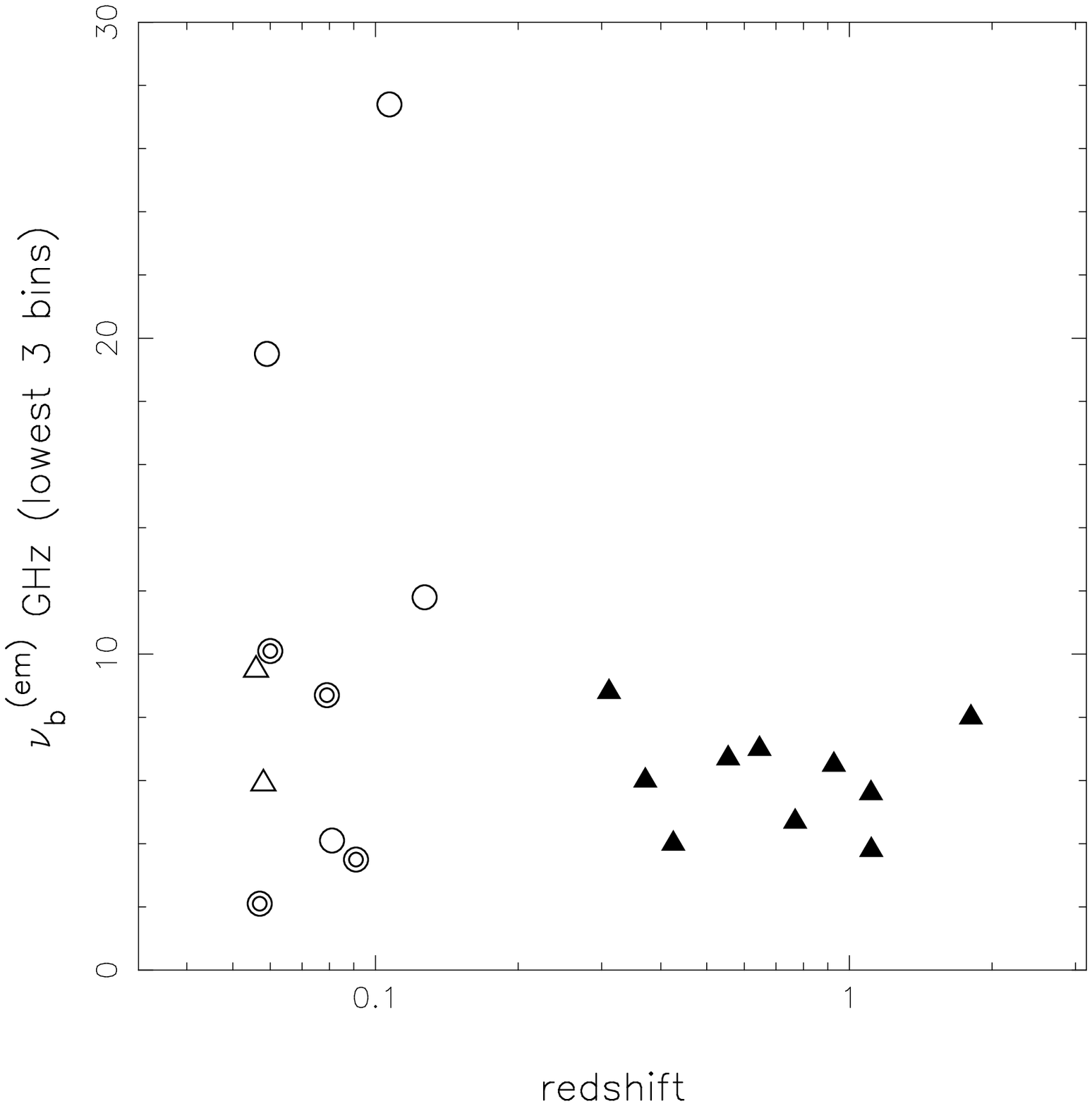, width
= 7cm, angle = -90}
\caption{(a) Observed spectra of lowest surface brightness
  regions. Mean of spectral indices in lowest three surface brightness
  bins on both jet and counter-jet side. (b) Associated break
  frequencies, using same theoretical spectrum as fig.~\ref{fig:break-hot}.
  Symbols as before.}

\label{fig:alpha-low-z}
\end{figure}

A similar analysis for the low surface-brightness structure shows a
correlation between $\alpha$ and $z$ for the quasars alone, but the
spectral indices for the galaxies are much more widely scattered than
those for the hotspots in the same objects (Fig.~\ref{fig:alpha-low-z}).
We again use model synchrotron-loss spectra to estimate the differences
between the two samples in the rest frame, with the same assumptions
employed for the hotspots in Section~4.1.  The result is shown in
Fig.~\ref{fig:alpha-low-z}b.  

The most striking feature is that two of the radio galaxies, 3C223.1 and
3C403, have very high values of $\nu_b$ (27 \& 20~GHz respectively).
The reason for this is clear from Fig.~\ref{fig:plots}: both sources
show essentially no spectral gradients ($\alpha$ averaged over lowest
three bins is $0.69$ and $0.77$).  Given that they are the two most
extreme examples of ``winged'' sources in our sample, we might naively
expect substantial synchrotron losses.  We have made additional
high-frequency observations of these sources and defer discussion of
this extremely puzzling result to a later paper.

Excluding these two peculiar sources, the mean values of $\nu_b$ are
similar for the galaxies and quasars and there is no evidence for a
correlation between $\nu_b$ and $z$. The errors on a given rest-frame
break frequency decrease with redshift, and this may account for the
lower dispersion for the quasars.  Increasing the injection index to
$\alpha = 0.7$ gives a qualitatively similar plot with larger values of
$\nu_b$.  With the exception of the ``winged'' sources, there is
therefore no evidence from our data alone for differences in the
rest-frame spectra of the low surface-brightness regions of the quasars
and radio galaxies.

It is clear, however, that the lowest surface-brightness bins are not in
the same parts of the lobes in two samples. In the radio galaxies, we
observe the steepest spectra close to the core, but the corresponding
regions in the quasars are usually below our surface-brightness limit.
As has already been remarked, the most striking difference between the
quasars (Fig.1 of Paper 1) and the radio galaxies
(Fig.~\ref{fig:images}) is that the lobes of all the radio galaxies
except 3C405 meet at the middle of the source, while most of the lobes
in the quasars stop well short of the core.  

The lack of detected emission toward the cores of the quasars occurs, at
least partly, because we observe the quasars at higher emitted
frequencies. Additionally, we might expect differences in rest-frame
spectra because the quasars have greater radio power and therefore,
probably, greater magnetic fields.  In either case, bridges should be
detected at lower frequencies.  As mentioned earlier in the context of
estimation of lobe volumes, Leahy \etal\ (1989)\nocite{lea89} have
indeed shown this to be the case for luminous radio galaxies and
quasars.  They also estimated values of $\nu_b$ for the
steepest-spectrum regions of the lobes (roughly equivalent to our low
surface-brightness bins), but using the hot-spot spectrum to define the
injection index and observing frequencies of 0.15 and 1.4~GHz.  For the
objects in common, Leahy \etal\ calculate break frequencies which are
lower than ours by a factor $\sim$ 4.  This provides some evidence that
lobes of the quasars have suffered more severe synchrotron losses than
those in the galaxies, but in view of the differences in assumed
injection spectrum and the large uncertainties, we do not regard it as
conclusive.  The fact that 3C405 shows the most extreme spectral
steepening close to the core, also suggests that synchrotron losses may
be more extreme in powerful sources.

\section{Orientation-dependent effects}

The new data presented here show no contradiction with the
interpretation of jet sidedness as an orientation effect; neither is
there any evidence that the morphologies or spectra of the hotspots and
associated emission share the sidedness of the jets. Black
(1992)\nocite{bla92b} found no tendency for the jet to be directed at
the smaller hot-spot or the hot-spot of higher surface brightness in the
full FRII sample of Black \etal\ (1992). Similarly, spectral asymmetries
in the present sample at high and low brightness levels are entirely
attributable to environmental or intrinsic effects.  In
contrast, the quasars described in Paper 1 and Bridle \etal\ 
(1994)\nocite{bri94} show asymmetries attributable to orientation: the
brightness of compact hotspots and the spectra of the hot-spot regions.
(They also show environmental/intrinsic effects: the spectra of the
extended regions and the lobe size).  In Paper 1, we concluded that
relativistic flow must persist beyond the hotspots and that the most
likely reasons for the spectral asymmetry are that a curved hot-spot
spectrum is Doppler shifted in frequency, or that we observe different
parts of the flow on the two sides as a result of Doppler enhancement
and suppression of the post-shock flow. There are two obvious
differences between the samples: the quasars are likely, on average, to
have axes closer to the line of sight and they are significantly more
powerful and distant. We address these points in turn.

The original sample of Black \etal\ (1992) was selected at a low frequency and
should, therefore, have a random distribution of orientations.  We have imposed
the additional criterion that at least one jet be detected.  Suppose that all
sources are intrinsically identical, with symmetrical, relativistic jets and
that we detect a fraction  $f$ of the sample with observed jet flux above some
limit.  These will then be within $\theta = \arccos (1-f) = 66^\circ$ of the
line of sight for the observed detection fraction of 60\% (this will, of
course, be an oversimplification because of sample variations).  Unified models
for powerful radio galaxies and quasars assume that they form a homogeneous
population, the broad-line region and nuclear continuum source being
obscured unless $\theta$ is less than some critical angle $\theta_c$.  A
source is classified as a quasar if $\theta < \theta_c$, otherwise as a
radio galaxy. Barthel (1989) estimated that $\theta_c \approx 45^\circ$
from the relative numbers of 3CR quasars and radio galaxies with $z >$ 0.5.
By this argument, roughly half of our galaxy sample is expected to lie in
the quasar orientation range.  Thus, although the galaxies should be
further from the line of sight on average, we should still expect to see
some orientation-dependent effects in their hotspots if flow speeds are
sufficiently fast.

If unified models also apply to our galaxy sample, then we can estimate
$\theta_c$ directly from the observed frequency of BLRG, which are thought
to be intrinsically weak quasars. Of the 26 FRII sources in the original
sample of Black \etal\ (1992), 6 (3C111, 227, 303, 382, 390.3 and 445) 
are classified
as BLRG, implying that that $\theta_c \approx 40^\circ$ (one other, 3C321, has
a broad H$\alpha$ line detected in polarized flux; Young \etal\
1996\nocite{you96}). This is likely to
be an underestimate since a substantial fraction of sources at this power
level show low-excitation or absorption-line optical spectra and cannot be
observed as BLRG at any orientation (Laing \etal\ 1994.\nocite{lai94-contrib}) We note
that the two BLRG in our galaxy sample (3C382 and 390.3) make good candidates
for being moderately close to the line of sight irrespective of
obscuration-based unified schemes: both have strong cores and 3C390.3 is a
possible superluminal (Alef \etal\ 1996).\nocite{ale96}

These arguments suggest that we might expect to see spectral asymmetries in
the high-brightness emission at least in the two BLRG, if not in other
sample members.  The fact that we do not, suggests that the
spectral differences between the quasar and galaxy samples 
reflect differences in power or redshift rather than orientation.  We
can identify three effects which could contribute to this difference, all
of them in the same sense:
\begin{enumerate}
\item the post-shock flow is likely to be slower in weaker jets;
\item the hot-spot spectra are observed at higher emitted frequency in the
more distant objects
\item the rest-frame spectra of the quasar hotspots show more curvature
than those of the galaxies (Section 4.1).
\end{enumerate}

\section{Summary and Conclusions}

We have obtained high quality radio images suitable for a two-frequency
spectral analysis for a sample of low-power (nearby) FRII radio galaxies
with detected jets.  The resolution is high enough to  distinguish hot-spot
and extended lobe material.

\subsection{Jet side and orientation}

\begin{enumerate}
\item Neither in this sample nor in a sample of
higher radio--power quasars (Paper 1) is there evidence of a
correlation between jet side and the spectral index of extended lobe
emission.
\item In the galaxy sample we do not detect the spectral asymmetry correlated
with jet side which we found in the hot-spot regions of the quasars.
\end{enumerate}

The spectral asymmetry in the high surface brightness regions of the
quasars is likely to be caused by material moving relativistically at
angles close to the line of sight.  The absence of any such asymmetry
in the radio galaxies could occur because they are, on average, closer
to the plane of the sky.  Alternatively, it may result from the
differences between the power and redshift distributions for the
samples.  We favour the latter explanation because we see no
asymmetries related to jet-side in the two broad-lined radio galaxies,
but this argument is not yet conclusive.

\subsection{Environmental/intrinsic effects}

\begin{enumerate}
\item In the low surface-brightness regions of both radio
galaxies and quasars, the larger lobe (by volume) has the flatter spectrum.
This asymmetry is consistent with differing rates of synchrotron loss or
adiabatic expansion resulting from an external density gradient (a more
detailed discussion is given in Paper 1).  
\item Two low-power radio galaxies with
large ``wings'' show remarkably uniform and flat spectra across
their entire structures. These sources will be the topic of a later paper.
\item The hot-spot spectra of the higher powered (more distant) 
sources are steeper at a given emitted frequency, so the observed
correlation between $\alpha$ and $z$ is not just the result of
redshifting a universal synchrotron-loss spectrum.
\end{enumerate}

\subsection*{ACKNOWLEDGEMENTS}

JDT, RAL and PAGS would like to thank the NRAO and Alan and Mary Bridle
for their hospitality. Thanks to Paul Alexander for {\sc anmap} software
used in production of theoretical spectra. We are grateful to W. van
Breugel, J. Burns, C. Carilli, J.P. Leahy, K. Roettiger and S. Spangler
for allowing us to use their data and images. JDT thanks the British
taxpayers for assistance in the form of a PPARC studentship.  This
research was partly supported by European Commission, TMR Programme,
Research Network Contract ERBFMRXCT96-0034 `CERES'. The NRAO is a
facility of the National Science Foundation, operated under cooperative
agreement by Associated Universities, Inc.

\end{document}